\documentclass[reprint, twocolumn]{aastex631}

\usepackage{amsmath,amssymb}
\usepackage{comment}

\newcommand{\Beq}{\begin{equation}\begin{aligned}}
\newcommand{\Eeq}{\end{aligned}\end{equation}}
\newcommand{\dd}{\mathrm{d}}

\begin{document}

\title{Primordial Black Hole Mergers as Probes of Dark Matter in Galactic Center} 

\author{Qianhang Ding}
\email{dingqh@ibs.re.kr}
\affiliation{Cosmology, Gravity and Astroparticle Physics Group, Center for Theoretical Physics of the Universe,
Institute for Basic Science (IBS), Daejeon, 34126, Korea}

\author{Minxi He}
\email{heminxi@ibs.re.kr}
\affiliation{Particle Theory and Cosmology Group, Center for Theoretical Physics of the Universe,
Institute for Basic Science (IBS), Daejeon, 34126, Korea}

\author{Volodymyr Takhistov}
\email{vtakhist@post.kek.jp}
\affiliation{International Center for Quantum-field Measurement Systems for Studies of the Universe and Particles (QUP), High Energy Accelerator Research Organization (KEK), 1-1 Oho, Tsukuba, Ibaraki 305-0801, Japan}
\affiliation{Theory Center, Institute of Particle and Nuclear Studies (IPNS), High Energy Accelerator Research Organization (KEK), Tsukuba 305-0801, Japan
}
\affiliation{Graduate University for Advanced Studies (SOKENDAI), \\
1-1 Oho, Tsukuba, Ibaraki 305-0801, Japan}
\affiliation{Kavli Institute for the Physics and Mathematics of the Universe (WPI), UTIAS, \\The University of Tokyo, Kashiwa, Chiba 277-8583, Japan}

\date{\today}

\begin{abstract}
Primordial black holes (PBHs) from the early Universe that can contribute to dark matter (DM) abundance have been linked to gravitational wave observations. Super-massive black holes (SMBHs) at the centers of galaxies are expected to modify distribution of DM in their vicinity, and can result in highly concentrated DM spikes.
We revisit PBH merger rates in the presence of DM spikes, tracking their history. We find novel peaked structure in the redshift-evolution of PBH merger rates at low redshifts around $z \sim 5$. These effects are generic and are present for distinct PBH mass functions and spike profiles, and also can be linked to peaked structure in 
redshift evolution of star formation rate.
Redshift evolution characteristics of PBH merger rates can be distinguished from astrophysical black hole contributions and observable with gravitational waves, enabling them to serve as probes of DM in galactic centers. 
\end{abstract}

\section{Introduction}

Primordial black holes (PBHs) could have formed in the early Universe and contribute to abundance of dark matter (DM)~(see e.g.~\citet{Sasaki:2018dmp,Carr:2020gox,Green:2020jor} for review). Recent gravitational wave (GW) detections by LIGO-Virgo-KAGRA (LVK) have been linked to stellar-mass black holes of primordial origin~(e.g.~\citet{Bird:2016dcv,Clesse:2016vqa,Sasaki:2016jop,Kashlinsky:2016sdv,Blinnikov:2016bxu}). Such PBHs can contribute to a substantial fraction of DM mass density $f_{\rm PBH} = \Omega_{\rm PBH}/\Omega_{\rm DM}$ as suggested by various complementary constraints including gas heating in dwarf galaxies~\citep{Lu:2020bmd,Takhistov:2021aqx,Takhistov:2021upb}, cosmic microwave background radiation~(e.g.~\citet{Ali-Haimoud:2016mbv,Poulin:2017bwe,Agius:2024ecw}), dwarf galaxy star dynamics~\citep{Brandt:2016aco,Koushiappas:2017chw,Graham:2023unf}, radio and X-ray observations~\citep{Inoue:2017csr, Manshanden:2018tze} as well as gravitational lensing~\citep{Zumalacarregui:2017qqd}. Current LVK observations imply $f_{\rm PBH} \lesssim \mathcal{O}(10^{-3})$ assuming that stellar-mass PBH mergers are responsible for the observed GW signals~(e.g.~\citet{Franciolini:2021tla}). However, the exact origin of these events remains uncertain. 

Variety of GW signatures associated with PBHs can originate at different stages of cosmic history and connect to distinct phenomena. Among them, PBH mergers can carry information about cosmic expansion~\citep{Ding:2022rpd, Ding:2023smy}, source stochastic GW background in different GW frequency bands~\citep{Mandic:2016lcn, Raidal:2017mfl, Cai:2021zxo}, 
and also serve as probes of primordial perturbations~\citep{Kimura:2021sqz, Wang:2022nml, Ding:2023smy}. PBHs can also source GWs not expected to originate from black holes of astrophysical origin, such as (sub-)solar mass black holes~\citep{Fuller:2017uyd,Takhistov:2017bpt,Takhistov:2017nmt,Bramante:2017ulk,Takhistov:2020vxs,Dasgupta:2020mqg,Wang:2021iwp,Sasaki:2021iuc,LIGOScientific:2022hai, Baumgarte:2024ouj,Crescimbeni:2024cwh, Yuan:2024yyo, Huang:2024wse, Crescimbeni:2024qrq}. 
The potential role of PBH mergers in GW observations as well as associated rich physics call for detailed investigation of their merger rates and evolution.

Observations definitively suggest presence of supermassive black hole (SMBH) Sgr A* residing 
in the Galactic Center of Milky Way~\citep{Ghez:1998ph,Eckart:1996zz}. More generally, 
SMBHs inhabit centers of galaxies~\citep{Volonteri:2021sfo}. Recently, observations by James Webb Space Telescope (JWST) of high-redshift active galactic nuclei (AGN) found prevalence of SMBHs~(e.g.~\citet{Matthee:2023utn,Ding:2023}). The presence of SMBHs can significantly modify distribution of DM in their vicinity. It has been argued~\citep{Gondolo:1999ef} that cold DM density around SMBHs can be dramatically enhanced forming a ``DM spike'' particularly when galactic halos follow a cuspy density profile as suggested by some N-body simulations~(e.g.~\citet{Navarro:2008kc,Stadel:2008pn}). 
Analyses based on general relativity~\citep{Sadeghian:2013laa, Speeney:2022ryg} further highlight significance of DM spike formation for observations. Recently, claims of DM spike detection based on SMBH binary orbital decay observations have been put forth~\citep{Chan:2024yht}, but require further scrutiny.
Further, DM spikes have also been studied in the context of intermediate-mass black holes and related GW observations~\citep{Zhao:2005zr,Bringmann:2009,Kavanagh:2020cfn,Aschersleben:2024xsb,Bertone:2024wbn}. More so, related formation of DM halos surrounding PBHs have been linked to novel signatures, including GWs~\citep{Coogan:2021uqv,Jangra:2023mqp} and gravitational lensing of fast radio bursts \citep{Oguri:2022fir}. Recently, it was demonstrated that GW lensing observations of PBHs in DM halos enable definitively probing scenarios of DM composed of combination of PBHs and particles~\citep{GilChoi:2023ahp}.

In this work, we establish novel and distinct features in PBH merger rate evolution driven by SMBH DM spikes. We demonstrate how PBH mergers can serve as intriguing probes of DM in galactic centers. While the presence of such DM spikes can significantly impact GW observations~\citep{Nishikawa:2017chy,Fakhry:2023ggw}, substantial uncertainties remain. To address this, we revisit the calculations of PBH mergers with SMBH DM spikes across variety of mass functions and redshift evolutions, providing a comprehensive understanding of their potential influence.

The paper is organized as follows. In Sec.~\ref{sec:dmdist} we introduce distribution of DM in galaxies and their centers, focusing on formation of DM spikes around SMBHs.
In Sec.~\ref{sec:smbhhalo} we discuss distribution of SMBH in DM halos.
In Sec.~\ref{sec:pbhmergers} we discuss PBH mergers originating from the early Universe as well as PBH mergers associated with DM spikes.  Then, in Sec.~\ref{sec:evolution} we analyze redshift evolution of PBH merger rates including effects of DM spike. 
Then, in Sec.~\ref{sec:evolutionspike} we discuss redshift evolution of DM spikes themselves.
In Sec.~\ref{sec:astrobh} we comment on astrophysical black holes with PBHs. We conclude in Sec.~\ref{sec:summary}.

\section{Dark matter in Galactic Center}
\label{sec:dmdist}

Structure formation $N$-body simulations typically favor a cuspy DM density distribution of the Galactic DM halo profile~\citep{Navarro:1995iw,Kravtsov:1997vm,Moore:1997sg}. This can be modeled using Navarro-Frenk-White (NFW) profile~\citep{Navarro:1995iw, Navarro:1996gj}
\begin{eqnarray}  \label{eq:dmprof}
\centering
\rho_{\rm DM}(r) = \dfrac{\rho_0}{(r / r_0)  (1+ r/r_0)^2}~,
\end{eqnarray}
with  $\rho_0 = 6.6\times 10^6\rm\ M_{\odot} / kpc^3$ and $r_{\rm 0} = 19.1\rm\ kpc$ for Milky Way.
More general DM distribution profiles can also be considered. However, mapping the DM distribution of the inner halo is challenging for observations. One can approximate the central galactic region DM density distribution using a power-law $ \rho(r) \simeq \rho_0 (r_0/r)^{\gamma} $ with index $ \gamma $ and halo parameters $ \rho_0 $ and $ r_0 $, which could be steeper than NFW. 

\begin{figure}[t] \centering
	\includegraphics[width=8cm]{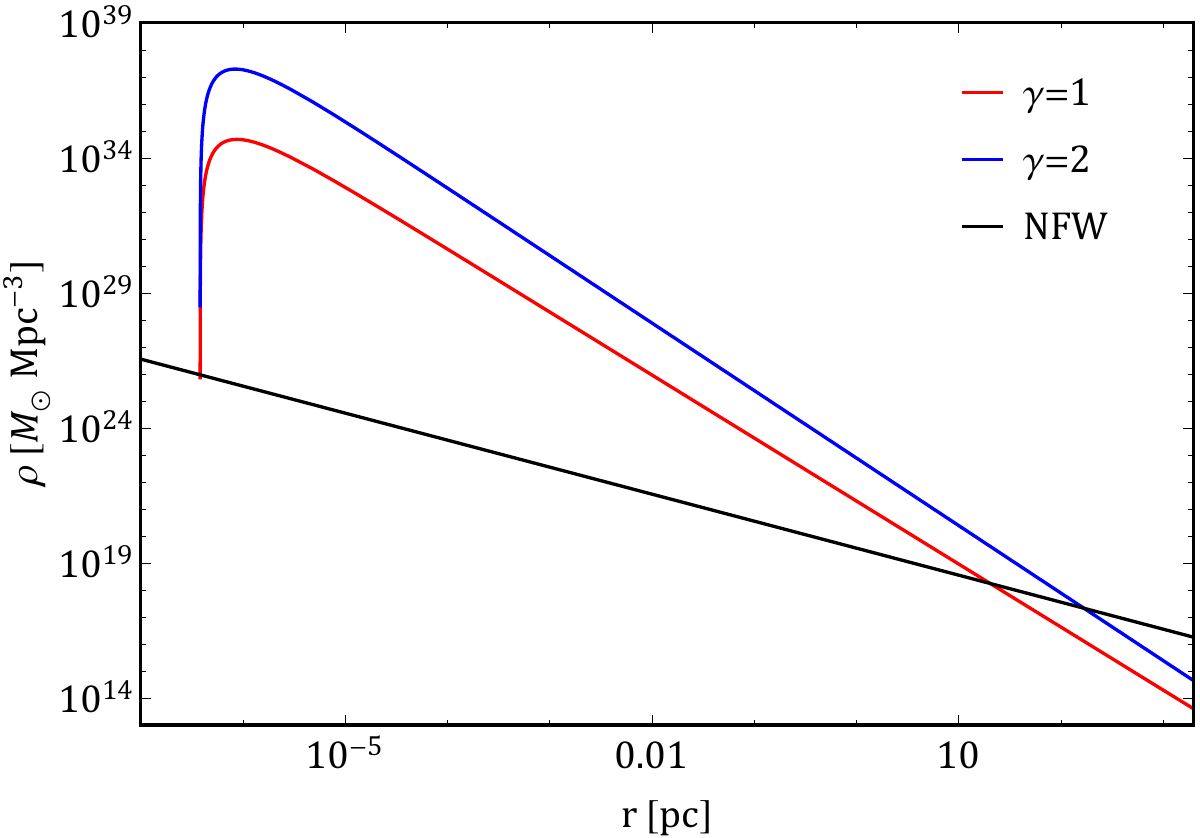}
\caption{
	Density distribution profile of DM spike around SMBH of mass $M_\mathrm{SMBH} = 10^6 M_\odot$ as a function of radial distance, considering spike profile power index of $\gamma = 1$ (red line) and $\gamma = 2$ (blue line). NFW profile (black line) considering DM halo parameters $\rho_0 = 3.7 \times 10^6 \, \rm M_\odot/ kpc^3$ and $r_0 = 9.6 \, \rm kpc$ is overlaid for reference.
	}
 \label{fig:spikeprofile}
\end{figure}

SMBHs residing in centers of galaxies~\citep{Volonteri:2021sfo} can dramatically affect surrounding distribution of DM. 
In the presence of a SMBH in a galactic center, it has been suggested that a dense spike of cold DM is expected to form due to SMBHs gravitational pull~\citep{Gondolo:1999ef}. On the other hand, 
the effects are negligible far from SMBH.
The density profile of the spike can be expressed as~\citep{Gondolo:1999ef}  
\begin{align}\label{eq:dm_spike}
    \rho_\mathrm{sp}(r) = \rho_R \left(1 - \frac{4 r_s}{r}\right)^3 \left(\frac{r_\mathrm{sp}}{r}\right)^{\gamma_\mathrm{sp}}~,
\end{align}
where $ \rho_R = \rho_0 (r_0/r_\mathrm{sp})^\gamma $ is the density at the boundary of the spike, $r_\mathrm{sp}$ is the radius of the DM spike, $ r_s = 2 G M_{\rm SMBH} $ is the Schwarzschild radius of the SMBH of mass $M_{\rm SMBH}$ at galactic center with $G$ being the gravitational constant, and $ \gamma_\mathrm{sp} = (9 - 2 \gamma)/(4 - \gamma) $ the power index of the spike. We do not consider here Kerr black holes, for which spike profile can be further enhanced depending on the BH spin~\citep{Ferrer:2017xwm}.  This description is approximately valid in the range of $4r_s < r < r_\mathrm{sp}$.
Throughout, we consider that DM halos follow NFW profile outside of DM spike region, for $r \gg r_{\rm sp}$.

In Fig.~\ref{fig:spikeprofile} we display two characteristic spike density profiles in the presence of a $ 10^6 M_\odot $ SMBH with $ \gamma =1 $ and $ \gamma =2 $, respectively. For such a SMBH we consider corresponding DM halo parameters $\rho_0 = 3.7 \times 10^6 \, \rm M_\odot/ kpc^3$ and $r_0 = 9.6 \, \rm kpc$. From Eq.~\eqref{eq:dm_spike}, the DM spike radius can be determined as 
\begin{equation}
r_\mathrm{sp}(\gamma, M_\mathrm{SMBH}) = \alpha_\gamma r_0 \Big(\dfrac{M_\mathrm{SMBH}}{\rho_0 r_0^3}\Big)^{1/(3-\gamma)}
\end{equation}
with normalization factor $ \alpha_\gamma $ for a given $ \gamma $.
Analyses grounded in general relativity~\citep{Sadeghian:2013laa, Speeney:2022ryg} underscore the significance of DM spike formation for observational studies.
They suggest DM spike profile should further extend the inner radius of DM spike closer to the SMBH, forming a greater spike density around inner radius.

The existence of DM spikes is still under debate. Recently, Ref.~\citet{Chan:2024yht} has claimed that observations of SMBH binary OJ 287 orbital decay are consistent with with dynamical friction originating from SMBH DM spikes with profile power index $\gamma_\mathrm{sp} \simeq 2.3$, corresponding to $\gamma \simeq 1$. However, further investigations are necessary. As we shall demonstrate, presence of such DM concentrations in the vicinity of galactic center SMBHs can carry significant implications for PBH mergers.

\section{Supermassive black holes in halos}
\label{sec:smbhhalo}

To estimate the effects of DM spikes on PBH mergers, 
we are interested in calculating PBH merger rates per DM halo hosting a central SMBH as a function of SMBH mass.
For this, we need to obtain relation between the mass of SMBH and the halo parameters $r_0$ and $\rho_0$ that will define the DM spike profile of Eq.~\eqref{eq:dm_spike}. 

We start from the relation between the mass of SMBH and DM halo velocity dispersion $\sigma$. Here, we follow the general approach of Ref.~\citet{Nishikawa:2017chy}, but also take into account effects of redshift $z$ evolution.
While the empirical correlation between SMBH masses and stellar velocity distribution $\sigma$ is typically considered, our consideration is also applicable to DM halos due to existence of a similar relation between velocity distribution in DM halos and SMBH mass~\citep{Larkin:2016dln}. 
Hence, the $M_\mathrm{SMBH}-\sigma$ relation is given\footnote{While Ref.~\citet{Robertson:2005fh} explored this for redshifts up to $z \sim 6$, at higher redshifts the $M_\mathrm{SMBH}-\sigma$ relation is uncertain. We have verified that our conclusions are not significantly affected if $M_\mathrm{SMBH}-\sigma$ relation is considered without significant redshift dependence.} by~\citep{Robertson:2005fh}
\begin{align}\label{eq:mass_sigma} \nonumber
    \log_{10}\Big(\dfrac{M_\mathrm{SMBH}}{M_\odot}\Big) =&~ a + b \log_{10}\Big(\dfrac{\sigma}{200 \, \mathrm{km} \, \mathrm{s}^{-1}}\Big) \\
    &- \xi \log_{10}(1 + z)~,
\end{align}
where parameters $a = 8.12 \pm 0.08$ and $b = 4.24 \pm 0.41$ have been empirically determined~\citep{Gultekin:2009qn}. This is also consistent with Ref.~\citet{Robertson:2005fh} that considered redshift evolution of the $M_{\rm SMBH}-\sigma$ relation, with
$\xi = 0.186$ being the coefficient for the redshift-dependent term.

Integrating over the radius NFW profile of Eq.~\eqref{eq:dmprof}, the spherical enclosed mass is 
\begin{equation}\label{eq:NFW_integration}
    M(r) = 4 \pi \rho_0 r_0^3 g \Big(\dfrac{r}{r_0}\Big)~,    
\end{equation}
where $g(y) = \log(1 + y) - y/(1 + y)$.
The NFW profile is taken to extend to virial radius $r_{\rm vir}$. 

The velocity dispersion $\sigma$ in an NFW halo corresponds to maximal circular velocity\footnote{That is, $v_{\rm c}(r)^2 = G M(r)/r$.} at radius $r_m = c_m r_0$, with $c_m = 2.16$, given by
\begin{align}\label{eq:vel_dis}
    \sigma^2 = \frac{G M(c_m r_0)}{c_m r_0} = \frac{4 \pi G \rho_0 r_0^2 g(c_m)}{c_m}~,
\end{align}
where $G$ is gravitational constant.
This enables expressing $ \sigma $ in Eq.~\eqref{eq:mass_sigma} is in terms of $ \rho_0 $ and $ r_0 $.
Then, for any given SMBH mass $M_\mathrm{SMBH}$ in Eq.~\eqref{eq:mass_sigma}, we can obtain its corresponding DM halo velocity dispersion $\sigma$ at different redshifts and apply it in 
Eq.~\eqref{eq:vel_dis}. This relates SMBH mass with $\rho_0$ and $r_0$.

\begin{figure}[t] \centering
	\includegraphics[width=8cm]{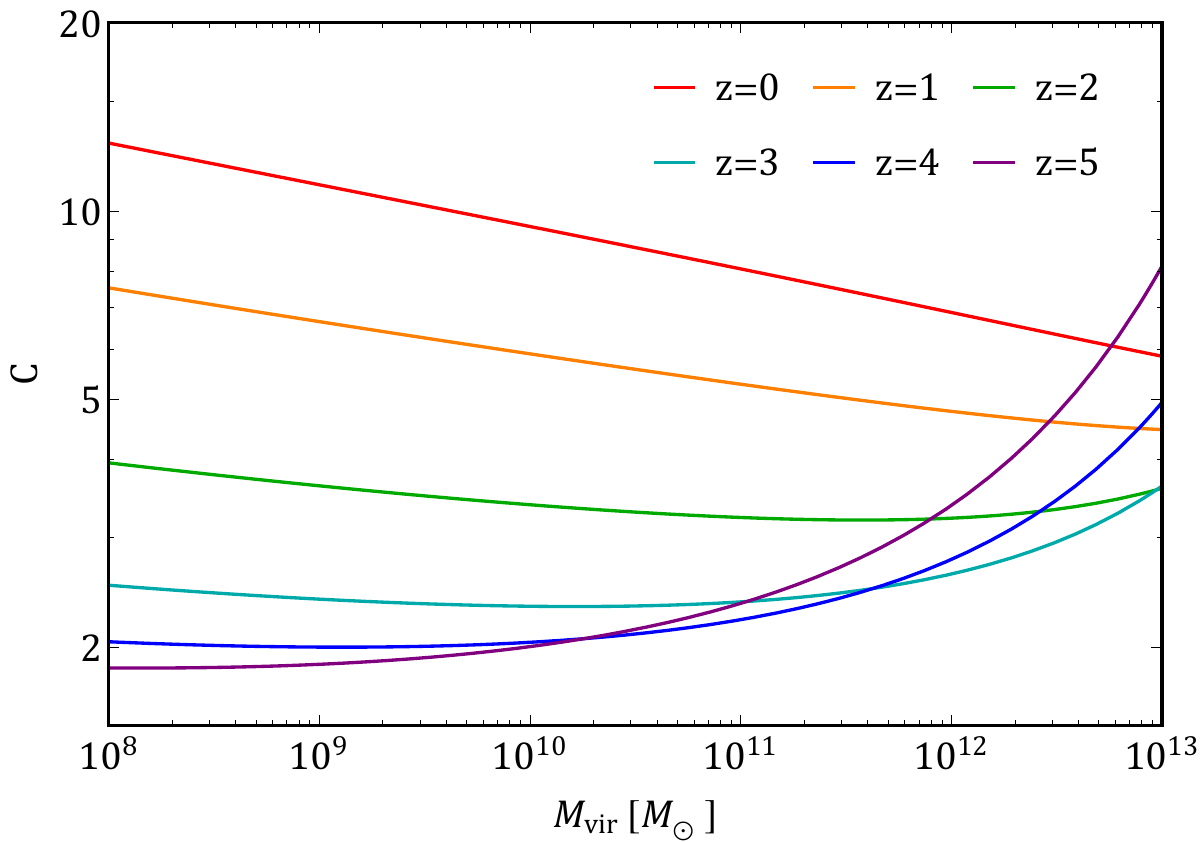}
	
	\caption{
		DM halo mass-concentration relation $ c (M_{\rm vir})$ considering different redshifts. Computed following method of Ref.~\citep{Prada:2011jf}.
	}
 \label{fig:concentration}
\end{figure}

In order to fix $\rho_0$ and $r_0$, we consider another independent relation between them. The virial mass of DM halo $M_\mathrm{vir}$ is the mass enclosed within the virial radius $r_\mathrm{vir}$. From definition
\begin{align}
    M_\mathrm{vir} &\equiv 200 \rho_\mathrm{crit} \left(\frac{4 \pi (c(M_\mathrm{vir}) r_0)^3}{3}\right) \label{eq:virial_mass_def} \\
    &= 4 \pi \rho_0 r_0^3 g(c(M_\mathrm{vir}))~, \label{eq:virial_mass_int}
\end{align}
 where $\rho_\mathrm{crit}$ is the critical energy density 
of the Universe and $c(M_\mathrm{vir}) \equiv r_\mathrm{vir}/r_0$ is the concentration parameter, which describes the concentration of DM mass in the halo. We compute numerically $c(M_{\rm vir})$ following procedure of Ref.~\citet{Prada:2011jf}. In Fig.~\ref{fig:concentration} we display mass-concentration relation at different redshifts considering the standard $\Lambda$CDM cosmological model with Planck 2018 parameters \citep{Planck:2018vyg}. The second line, Eq.~\eqref{eq:virial_mass_int}, is obtained by integration over the NFW density profile within $ r_{\rm vir} $ in Eq.~\eqref{eq:NFW_integration}. Then, we construct independent relation between $r_0$ and $\rho_0$ by combining Eqs.~\eqref{eq:virial_mass_def} and \eqref{eq:virial_mass_int} with mass-concentration relation. This relation with the one determined from Eqs.~\eqref{eq:mass_sigma} and \eqref{eq:vel_dis} determines $\rho_0$ and $r_0$. Combined together with the mass of DM halo $M_\mathrm{vir}$ from Eq.~\eqref{eq:virial_mass_def}, this allows us to construct   $M_\mathrm{vir}$ as a function of $M_{\rm SMBH}$ at different redshifts. 

To construct the SMBH mass function we also require DM halo mass function $\dd n/\dd M_\mathrm{vir}$, which can be calculated as~\citep{2012MNRAS.423.3018P}
\begin{align} \label{eq:halofun}
    \frac{\dd n}{\dd M_\mathrm{vir}} = f(\sigma_M) \frac{\rho_m}{M_\mathrm{vir}} \frac{\dd \log (\sigma_M^{-1})}{\dd M_\mathrm{vir}}~,
\end{align}
where $\rho_m(z) = \rho_{m,0} (1+z)^3$ is the cosmological matter density that depends on redshift and $\rho_{m,0} = 39.7 \, M_\odot /\rm kpc^3 $ is the matter density at present. Here, $\sigma_M$ is the linear root-mean-square fluctuation of density field on the scale $M_\mathrm{vir}$, and can be calculated from a power-spectrum of density fluctuations $P(k,z)$ as
\begin{equation}\label{eq:sigma_m}
    \sigma_M^2(M_\mathrm{vir}, z) = \frac{1}{2 \pi^2} \int_0^\infty P(k, z) W^2(k, M_\mathrm{vir}) k^2 \dd k~,
\end{equation}
 and the power-spectrum $P(k,z)$ is calculated via correlation function of matter density contrast at different redshifts as
\begin{align}
    \langle \delta(\Vec{x}) \delta(\Vec{x}) \rangle = \int_0^\infty \frac{P(k)}{2 \pi^2} k^2 \dd k~,
\end{align}
with matter density contrast defined as $\delta(\Vec{x}) \equiv (\rho(\Vec{x}) - \bar{\rho})/\bar{\rho}$. This can be calculated considering standard cosmological $\Lambda$CDM model with parameters $(\Omega_m, \Omega_\Lambda, \Omega_b, n_s, h, \sigma_8) = (0.27,0.73,0.0469,0.95,0.70,0.82) $, where $\Omega_X \equiv \rho_X/\rho_\mathrm{crit}$ is abundance in terms of critical density $\rho_{\rm crit} = 2.78 \times 10^{11}h^2 M_{\odot}$Mpc$^{-3}$ with $X = (m, \Lambda, b)$, $n_s$ is the spectral index of the primordial power-spectrum, $h \equiv H_0/100 \, \rm km \, s^{-1} Mpc^{-1}$ and $\sigma_8$ is the root-mean-square amplitude of linear mass fluctuations in spheres of $8 h^{-1} \rm Mpc$. The top-hat filter function $W(k, M_\mathrm{vir})$ is defined as
\begin{equation}
    W(k, M_\mathrm{vir}) = 
    \begin{cases}
        1 & 0 < k < 1/r_\mathrm{vir}~,\\
        0 & \text{otherwise}~.
    \end{cases}
\end{equation}
Then, $f(\sigma_M)$ function accounts for the geometry of the collapsing overdense regions and can be estimated as
\begin{equation}
f(\sigma_M) = A \left(1 + \Big(\frac{\sigma_M}{b}\right)^{-a}\Big) \exp \left(- \frac{c}{\sigma_M^2}\right)~,
\end{equation}
where we consider a spherical collapse with the parameters $A = 0.213$, $a = 1.8$, $b = 1.85$, and $c = 1.57$ as in Ref.~\citet{Tinker:2008ff}. We calculate $\sigma_M(z)$ following approximate semi-analytic treatment of Ref.~\citet{Klypin:2011}.
 
\begin{figure}[t] \centering
	\includegraphics[width=8cm]{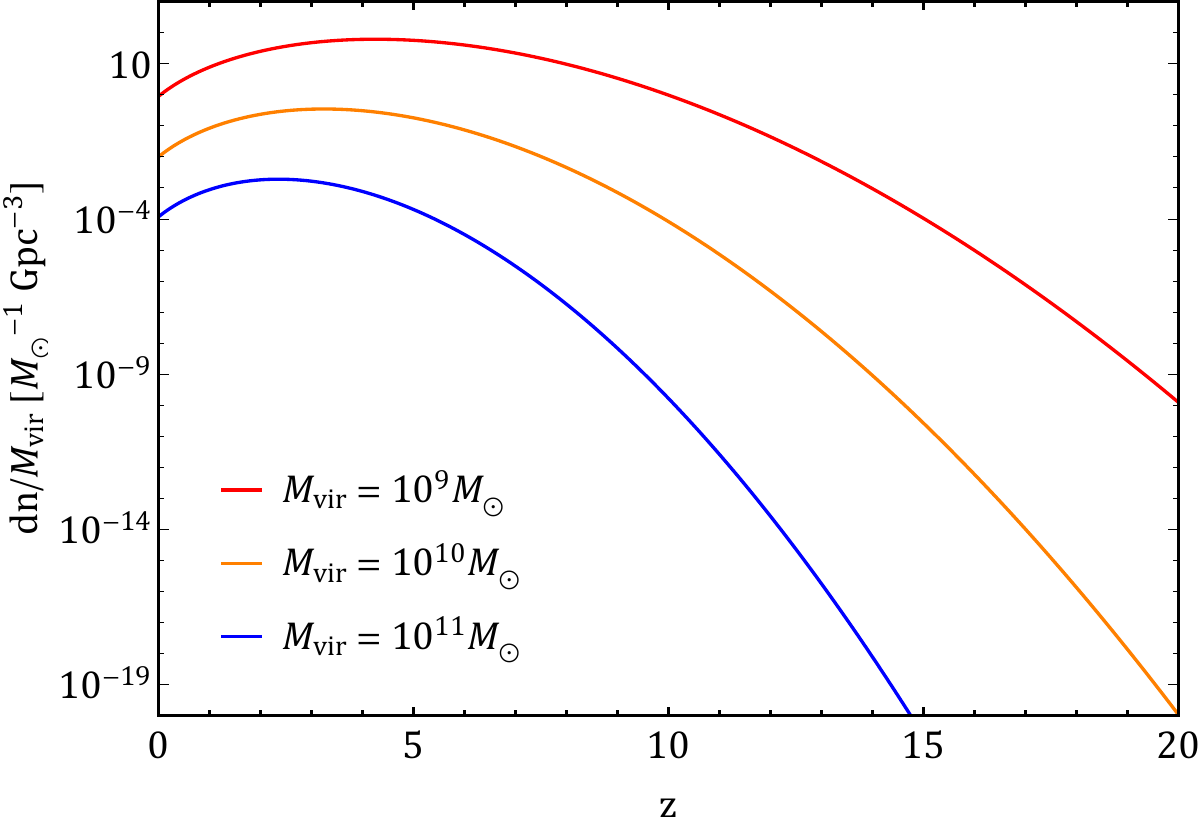}
	\caption{
        Redshift evolution of the differential number density of DM halos for different halo masses $M_\mathrm{vir}$, derived from Eq.~\eqref{eq:halofun}.  
	}
 \label{fig:halo_redshift}
\end{figure}

In Fig.~\ref{fig:halo_redshift} we display differential number density of DM halos for various halo masses, from Eq.~\eqref{eq:halofun}. We observe characteristic peaked structure around redshift $z \sim $~few. In Sec.~\ref{app:sfr} we reconstruct star formation rate (SFR) redshift evolution that is in agreement with observations. We confirm that SFR peaked structure around redshift of $z \sim$~few originates from DM halo mass function in this case as well. As we will show, these effects can also significantly impact redshift evolution of PBH merger rates.

Combining results, we obtain theoretical prediction for SMBH mass function distribution given by
\begin{align}\label{eq:SMBH_massfun}\nonumber
    \frac{\dd n}{\dd M_\mathrm{SMBH}} &= \frac{\dd n}{\dd M_\mathrm{vir}} \frac{\dd M_\mathrm{vir}}{\dd M_\mathrm{SMBH}} \\
    &= f(\sigma_M) \frac{\rho_m}{M_\mathrm{vir}} \frac{\dd \log (\sigma_M^{-1})}{\dd M_\mathrm{vir}} \frac{\dd M_\mathrm{vir}}{\dd M_\mathrm{SMBH}}~.
\end{align}
Here, the redshift dependence of SMBH mass function comes from redshift behavior of various parameters, including $\sigma_M^2(M_\mathrm{vir}, z)$ in Eq.~\eqref{eq:sigma_m}, $\rho_m(z)$, and $M_\mathrm{SMBH} - M_\mathrm{vir}$ relation that determining $\dd M_{\rm vir}/\dd M_{\rm SMBH}$.  
Using obtained $M_\mathrm{SMBH} - M_\mathrm{vir}$ relation at different redshifts,
we can eliminate in Eq.~\eqref{eq:SMBH_massfun} dependency on $M_{\rm vir}$ for SMBH mass $M_{\rm SMBH}$. We assume the spherical halo collapse model and standard $\Lambda$CDM cosmology for relevant computations. 

\begin{figure}[t] \centering
	\includegraphics[width=8cm]{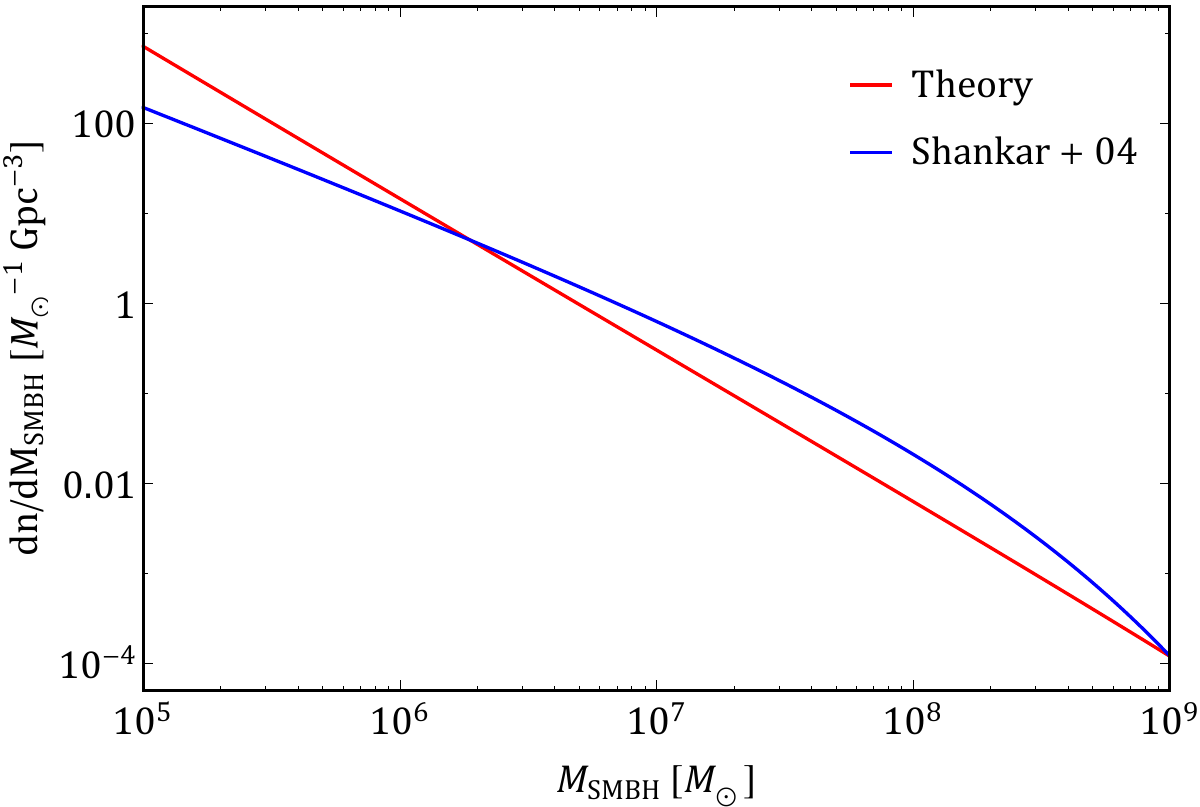}
	\includegraphics[width=8cm]{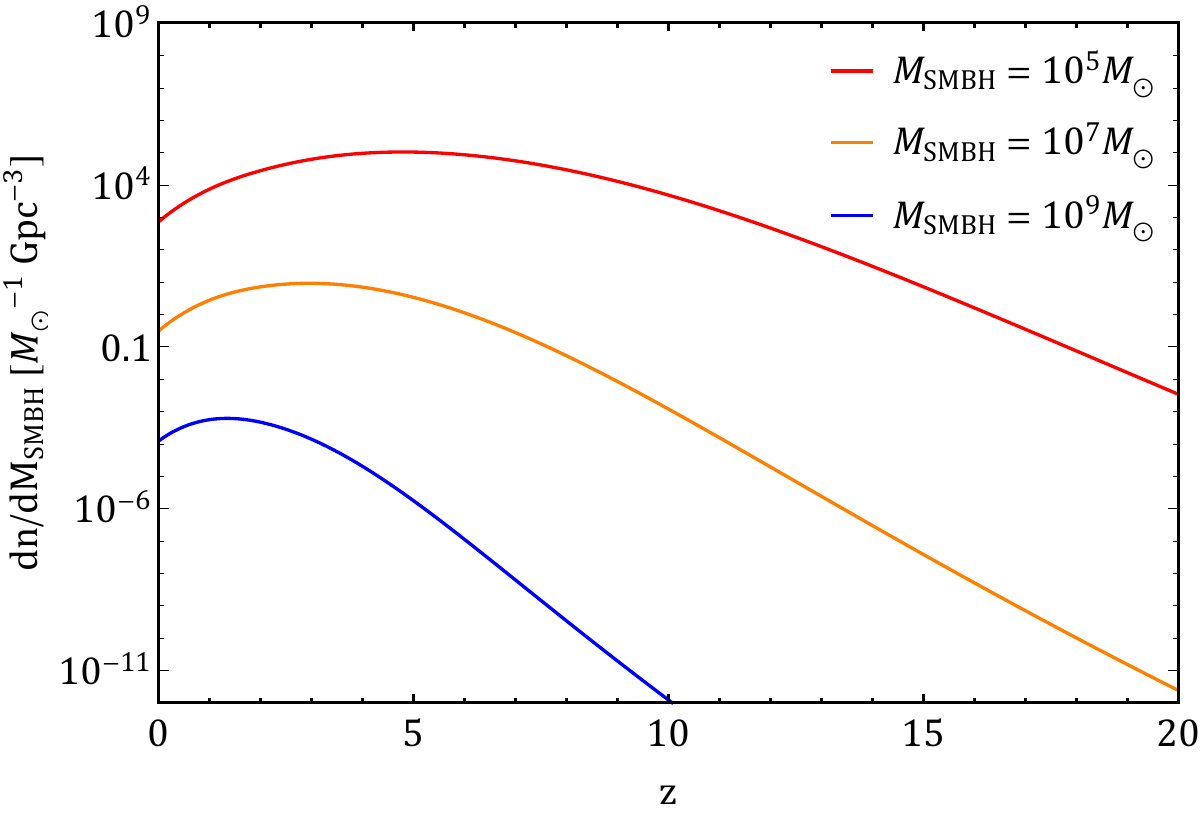}
	\caption{
	[Top] SMBH mass function distribution from theoretical calculation (red line) described by Eq.~\eqref{eq:SMBH_massfun} and empirical fit to kinematic and photometric data of quasars and AGNs (blue line) \citep{Shankar:2004ir}. 
        [Bottom] Redshift evolution of SMBH differential number density for different SMBH masses $M_\mathrm{SMBH}$.  
	}
 \label{fig:SMBH_distribution}
\end{figure}

In the upper panel of
Fig.~\ref{fig:SMBH_distribution}, we
compare theoretical predictions described by Eq.~\eqref{eq:SMBH_massfun} with empirical SMBH mass function based on kinematic and photometric data of quasars and active
galactic nuclei (AGNs) in Ref.~\citet{Shankar:2004ir}. We observe good qualitative agreement. 
In the lower panel of Fig.~\ref{fig:SMBH_distribution} we display the redshift behavior of SMBH differential number density, and observe peaked features for different SMBH masses that affects the redshift evolution of PBH merger rates in the following discussion. We note that the SMBH mass-function evolution we consider assumes existence of SMBHs up to higher redshifts $z \sim \mathcal{O}(10)$. Hundreds of quasars have already been discovered at redshifts $z > 6$ with different considerations for their formation~(see e.g.~\citet{Fan:2023} for review), and recent JWST observations indicate accreting SMBHs at $z > 8.5$~\citep{Larson:2023}.

\section{Primordial black hole mergers}
\label{sec:pbhmergers}

PBH mergers can arise from distinct formation channels across cosmic history, which contribute to the total merger rate. After PBH formation, PBHs that are typically considered to be initially Poisson-distributed in the early Universe form binaries through multi-body interactions and capture at high redshifts. These merging PBH binaries, which we call early PBH mergers (EPM), predominantly form before the large-scale structure of the Universe develops, although their mergers can happen much later. On the other hand, as cosmic structure forms, PBHs contributing to DM abundance, along with the rest of DM, cluster into DM halos. Then, halo PBH mergers can take place in dark matter halos, where the local PBH and DM densities are significantly higher than the cosmic average. These include PBH binaries formed within halos or from late-time clustering of PBHs, resulting in enhanced merger rates due to the denser environment.  For detailed recent investigation of PBH mergers in DM halos see e.g. Ref.~\citet{Aljaf:2024fru}.

Here, we focus on another distinct contributing channel to PBH mergers originating from centers of galactic DM halos around SMBHs in high DM density spikes. 

\subsection{Early binary mergers}

Formation of EPM binaries in the early Universe and their merger GW signals at both low and high redshifts have been extensively studied~\citep{Sasaki:2016jop,
Ali-Haimoud:2017rtz, Raidal:2018bbj, Ng:2022agi}. Such PBH binaries would have a small semi-axis and a large eccentricity, which cause the merger timescale much shorter than binary formation and Hubble timescale~\citep{Sasaki:2016jop}. The resulting differential merger rate per comoving volume is
\begin{align}\label{eq:background_merger}\nonumber
    \frac{\dd R}{\dd m_1 \dd m_2} =&~ \frac{1.6 \times 10^6}{\mathrm{Gpc}^3 \mathrm{yr}} f_\mathrm{PBH}^{\frac{53}{37}} \left(\frac{t(z)}{t_0}\right)^{-\frac{34}{37}} \eta^{-\frac{34}{37}} \left(\frac{M}{M_\odot}\right)^{-\frac{32}{37}}\\
    &~\times  S(M, f_\mathrm{PBH}) \psi(m_1) \psi(m_2)~,
\end{align}
where $ \psi (m) $ is PBH mass distribution, $ M=m_1 +m_2 $ is the total mass of the two PBHs with masses $ m_1 $ and $ m_2 $, $ \eta=m_1 m_2/M^2 $ the symmetric mass fraction.  
The suppression factor $ S(M, f_\mathrm{PBH}) $ stands for the interaction between PBH binaries with environment that could disrupt the binaries, see e.g. Ref.~\citet{Hutsi:2020sol}, however there is uncertainty on these effects.
Here, we employ approximate treatment of suppression factor\footnote{This slightly underestimates the rate when $f_{\rm PBH} \ll \sigma_M$.} from~\citet{Ali-Haimoud:2017rtz, Raidal:2018bbj} as
$ S(M, f_\mathrm{PBH}) = (1 + \sigma_b^2/f_{\rm PBH}^2)^{-21/74}$, where $\sigma_b = 1.4 \times 10^{-2}$ is the rescaled variance of matter density perturbations at the time the binary is formed. 
The results from Eq.~\eqref{eq:background_merger} consider that PBHs after formation follow Poisson distribution.

The PBH mass distribution $ \psi (m) $ sensitively depends on the models of PBH formation. We consider two characteristic examples, a monochromatic type and a log-normal type. 
Here, $\psi(m)$ is defined as 
\begin{equation}
   \psi(m) = \frac{1}{\rho_\mathrm{DM}}\frac{\dd \rho_\mathrm{PBH}}{\dd m}~,
\end{equation}
where $\rho_{\rm DM}$ is the DM density. The mass function is normalized as $\int \psi(m) \dd m = f_\mathrm{PBH}$. 
Monochromatic 
PBH mass function can be viewed as Dirac delta function centered around characteristic PBH mass $m_c$. The log-normal PBH mass function can be expressed as 
\begin{align}
    \psi(m) = \frac{f_\mathrm{PBH}}{\sqrt{2 \pi} \sigma m} \exp \left(-\frac{\log(m/m_c)^2}{2 \sigma^2}\right)~,
\end{align}
where $\sigma$ is the width of the mass distribution. 

\subsection{Mergers in halos} 

In addition to EPM, PBH binaries can also form and merge in overdense regions through interactions, such as scattering and gravitational bremsstrahlung emission~\citep{Bird:2016dcv}. 
During encounter of two PBHs if sufficient amount of energy is lost due to GWs they can form a gravitationally bound system.
The associated merger rate can be approximated as the capture rate of PBHs in overdense regions. 

DM halos, where the DM density is several orders of magnitude larger than the average cosmological DM density in the Universe, constitute a favorable environment for such PBH binary formation.  The merger rate of PBHs in a single halo with a virial radius $r_{\rm vir}$, considering a monochromatic PBH mass function for PBHs of mass $M_{\rm PBH}$, can be expressed as 
\begin{align}\label{eq:merger_in_halo}
    N_{\rm halo} = \int_{}^{r_{\rm vir}} \frac{1}{2} \left(\frac{f_\mathrm{PBH} \rho_{\rm DM}(r)}{M_\mathrm{PBH}}\right)^2 \sigma_m(r) v_\mathrm{rel}(r) \dd^3 r~,
\end{align}
where $\rho_{\rm DM}$ is the DM halo density profile such as NFW of Eq.~\eqref{eq:dmprof}, $v_\mathrm{rel}$ is the relative velocity between PBHs and $\sigma_m(r)$ is the two-body PBH scattering cross-section for GW emission, which is~\citep{Mouri:2002mc}
\begin{align}\label{eq:cross_section}
    \sigma_m(r) = 1.4 \times 10^{-14} \left(\frac{M_\mathrm{PBH}}{30 \, M_\odot}\right)^2 \left(\frac{v_\mathrm{rel}(r)}{200 \, \mathrm{km} \, \mathrm{s}^{-1}}\right)^{-\frac{18}{7}} \mathrm{pc}^2~.
\end{align} 
The total PBH DM halo merger rate can then be found from integrating Eq.~\eqref{eq:merger_in_halo} over the distribution of DM halos. Note that in principle in Eq.~\eqref{eq:merger_in_halo} $\sigma_m(r) v_{\rm rel}(r)$ corresponds to averaged value $\langle\sigma_m v_{\rm rel} \rangle$ over relative PBH DM velocity distribution in halo that can be approximated by a Maxwell-Boltzmann distribution (e.g.~\citet{Bird:2016dcv}). On the other hand, for PBH merger rates in DM spike the relevant relative velocity can be approximated by circular velocity around SMBHs.

The PBH DM halo merger rate calculation can also be readily applied to PBHs with an extended mass-function (e.g. log-normal), through
\begin{align}\label{eq:extended_merger}\nonumber
    N_\mathrm{halo} =&~ \int \int \int_{}^{r_{\rm vir}} \frac{1}{2} \left(\frac{\psi(m_1) \rho_\mathrm{DM}(r)}{m_1}\right)\left(\frac{\psi(m_2) \rho_\mathrm{DM}(r)}{m_2}\right)\\
    &~\times \sigma_m(m_1,m_2,r) v_\mathrm{rel}(r) \, \dd^3 r \, \dd m_1 \, \dd m_2~.
\end{align}
Here, the process cross-section accounts for two distinct contributing possible PBH masses $m_1$ and $m_2$ \citep{Mouri:2002mc},
\begin{align}
    \sigma_m(m_1,m_2,r) =&~ 2 \pi \left(\frac{85 \pi}{6 \sqrt{2}}\right)^{2/7} G^2 \\
    &~\times \frac{(m_1 + m_2)^{10/7} m_1^{2/7} m_2^{2/7}}{
    v_\mathrm{rel}(r)^{18/7}}~. \notag
\end{align}

Then, total PBH mergers in DM halo can be evaluated via integrating over halo mass function in Eq.~\eqref{eq:halofun} as
\begin{align}
    R_\mathrm{halo} = \int_{M_\mathrm{vir, min}}^{M_\mathrm{vir, max}} N_\mathrm{halo} \frac{\dd n}{\dd M_\mathrm{vir}} \dd M_\mathrm{vir}~,
\end{align}  
PBH mergers in DM halos have been extensively studied with analyses finding that DM halo merger rates are typically significantly subdominant compared to that of EPM. For instance, assuming $f_{\rm PBH} = 1$, the merger rate of $30 \, M_\odot$ PBHs in DM halos is around $\mathcal{O}(10) \, \rm Gpc^{-3} \, yr^{-1}$~\citep{Bird:2016dcv, Ali-Haimoud:2017rtz, Fakhry:2020plg}, while EPM results in a merger rate of $\mathcal{O}(10^5) \, \rm Gpc^{-3} \, yr^{-1}$ estimated from Eq.~\eqref{eq:background_merger}. 

\subsection{Mergers in galactic center spikes}\label{sec:merger_spike}

\begin{figure}[tbp] \centering
	\includegraphics[width=8cm]{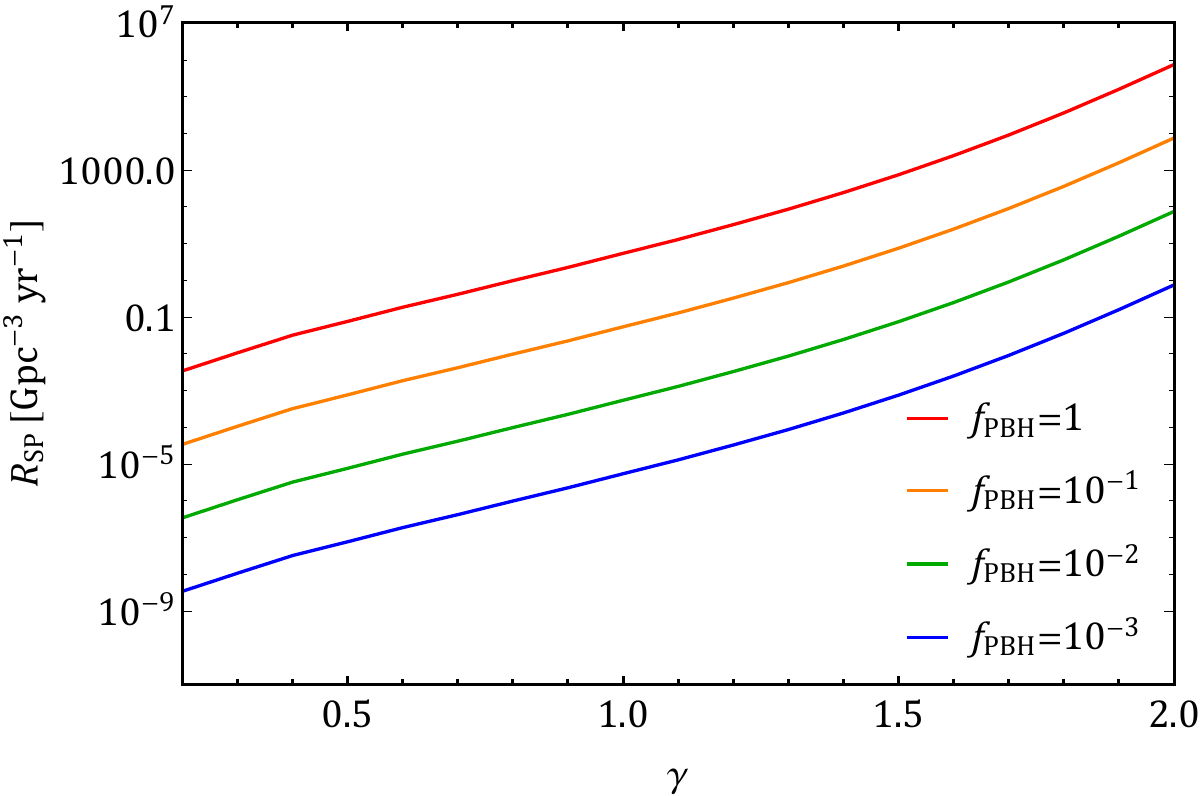}
	\caption{
Dependence of total PBH merger rate from DM spikes on the DM spike profile index $\gamma$ for different $f_{\rm PBH}$. We consider PBH mass of $M_\mathrm{PBH} = 30 \, M_\odot$, as well as lower bound on SMBH masses of $M_\mathrm{SMBH,min} = 10^5 \, M_\odot$ and upper bound of $M_\mathrm{SMBH,max} = 10^9 \, M_\odot$, respectively.
	}
        \label{fig:merger_gamma}
\end{figure}

In galactic centers, presence of extremely high DM spike overdensities can significantly enhance the merger rate of PBHs that can be comparable to or even exceed the  EPM rate at present time depending on spike density profile~\citep{Nishikawa:2017chy,Fakhry:2023ggw}.
Here, we revisit PBH merger rate contributions from DM spikes. As we will demonstrate in Sec.~\ref{sec:evolution}, this can significantly affect the PBH merger rate redshift evolution in the late Universe with novel emerging features that could be observable in GW experiments.

Analogously to DM halo PBH mergers, merger rate of PBH binaries in a DM spike can be evaluated by considering the binary formation rate of PBHs from their interactions. 
Following Eq.~\eqref{eq:merger_in_halo}, the DM spike merger rate of PBHs with a monochromatic mass function can be expressed as  
\begin{align}\label{eq:merger_in_spike}
    N_\mathrm{sp} = \int_{4 r_s}^{r_\mathrm{sp}} \frac{1}{2} \left(\frac{f_\mathrm{PBH} \rho_\mathrm{sp}(r)}{M_\mathrm{PBH}}\right)^2 \sigma_m(r) v_\mathrm{rel}(r) \dd^3 r~,
\end{align}
where $\rho_{\rm sp}$ is the spike DM density profile described by Eq.~\eqref{eq:dm_spike}, and the radial integration is over the spike contributions $4r_s < r < r_\mathrm{sp}$.
For the relative velocity, we use the circular velocity around SMBH, as in analysis of Ref.~\citet{Nishikawa:2017chy},
\begin{equation} \label{eq:vrel}
    v_\mathrm{rel} = \Big(\dfrac{G M_\mathrm{SMBH}}{r}\Big)^{1/2}
\end{equation} 
for each considered radius. We note that some fraction of PBHs contributing to DM spike might have sizable orbits within DM halo and hence spending only a fraction of their time in SMBH vicinity. In such case, relevant relative velocity can be distinct from that of Eq.~\eqref{eq:vrel}. Considering velocity scaling of cross-section in Eq.~\eqref{eq:cross_section}, reduced $v_{\rm rel}$ can result in enhanced PBH merger rate. This highlights the need for further detailed analyses based on simulations to determine PBH behavior in such environments.

Here, we focus on merger rate contributions from simplest two-body capture without including other channels, such as three-body interactions. The PBH merger rate from two-body captures are typically dominant, significantly exceeding three-body interactions for a small value of $f_\mathrm{PBH}$~\citep{Sadeghian:2013laa, Franciolini:2022ewd}, which is our main interest with multiple observations constraining $f_\mathrm{PBH} \ll 1$ for stellar-mass PBHs. For an extended PBH mass-function, we calculate the PBH merger rate in analogy with Eq.~\eqref{eq:extended_merger} and considering radial limits of integration $4r_s < r < r_\mathrm{sp}$ and DM density in spike $\rho_{\rm sp}$.

In order to obtain the total PBH merger rate from DM spikes per comoving volume, we need to account for the PBH merger contributions from all DM spikes. The PBH merger rate in DM spike $N_\mathrm{sp}$ depends on the profile of DM spike, which is determined by SMBH mass. Hence, contributions from all DM spike can be described by the mass-function of SMBHs. Then, the total PBH merger rate in DM spikes per volume can be found from
\begin{align}\label{eq:total_merger_spike}
R_{\rm SP} = \int_{M_\mathrm{SMBH, min}}^{M_\mathrm{SMBH, max}} N_\mathrm{sp}(M_\mathrm{SMBH}) \frac{\dd n}{\dd M_\mathrm{SMBH}} \dd M_\mathrm{SMBH}~,
\end{align}
where $M_\mathrm{SMBH,min}$ and $M_\mathrm{SMBH,max}$ are the minimal and the maximal SMBH masses we consider, respectively. The mass range of SMBHs in galactic centers is assumed to be between $M_\mathrm{SMBH,min} = 10^5 - 10^6 \, M_\odot$ and $M_\mathrm{SMBH, max} = 10^9 - 10^{10} \, M_\odot$, as in typical mass-functions. From Eq.~\eqref{eq:total_merger_spike}, a larger SMBH number density would increase the total PBH merger rates. 
The number density of lighter SMBHs is larger, as shown in Fig.~\ref{fig:SMBH_distribution},and hence generally $ R_{\rm SP} $ is more sensitive to the lower limit of SMBH masses that we consider to be $M_\mathrm{SMBH, min} = 10^5 \, M_\odot$ unless stated otherwise. We also typically consider the upper bound on SMBH mass of $M_\mathrm{SMBH, max} = 10^9 \, M_\odot$ in our calculations.

In~Fig.~\ref{fig:merger_gamma} we display the total PBH merger rate from DM spikes from Eq.~\eqref{eq:total_merger_spike} for different $f_\mathrm{PBH}$ and spike profiles described by index $\gamma$, considering
monochromatic PBH mass-spectrum with mass $M_{\rm PBH} = 30 \, M_\odot$. This clearly demonstrates that when DM spike profile index $\gamma$ becomes larger, the PBHs merger rate in DM spike is enhanced. We also find that the total PBH DM spike merger rate behaves as $R_{\rm SP} \sim f_\mathrm{PBH}^2$, which can be understood from Eq.~\eqref{eq:merger_in_spike}. In particular, when $\gamma = 2$ the PBH merger rate from the DM spike contributions are seen to approach $\sim 10^6 \, \mathrm{Gpc}^{-3} \, \mathrm{yr}^{-1}$, and for steeper profiles described by larger $\gamma$ power can be comparable with that from EPMs as we find from Eq.~\eqref{eq:background_merger}.
This indicates that PBH merger rate in DM spikes could significantly modify the total PBH merger rate evolution in the late Universe, and hence PBH mergers can serve as probes of DM concentration in galactic centers.

\section{Merger rate evolution}
\label{sec:evolution}

The formation of DM spikes begins at high redshifts. Thus, redshift evolution of the PBH merger rates can be accordingly modified. As we show, these contributions can dominate PBH merger rates and novel features that depend on redshift evolution emerge.
These effects offer new insights into interpreting PBH merger signals detected by GWs observations.

\subsection{Merger rate redshift evolution peaks}

Several factors affect the total merger rate redshift evolution of PBHs as described in Eq.~\eqref{eq:total_merger_spike}, the PBH merger rate in each DM spike $N_\mathrm{sp}$ and the differential number density of SMBHs $\dd n/\dd M_\mathrm{SMBH}$. 
The PBH merger rate in a DM spike $N_\mathrm{sp}$ depends on the DM spike profile, which is determined by the SMBH mass.   

Differential number density of SMBHs $\dd n/\dd M_\mathrm{SMBH}$ could play an important role in the redshift evolution of PBH merger rates. 
As shown in Eq.~\eqref{eq:SMBH_massfun}, the higher matter density $\rho_m$ and rapid SMBH formation can increase $\dd n/\dd M_\mathrm{SMBH}$ at higher redshifts, while a decreasing $f(\sigma_M)$ would suppress $\dd n/\dd M_\mathrm{SMBH}$. The redshift-dependent behavior of $\dd n/\dd M_\mathrm{SMBH}$ for different SMBH masses is shown in the lower panel of Fig.~\ref{fig:SMBH_distribution}.
We observe a peaked feature in $\dd n/\dd M_\mathrm{SMBH}$, with the peak location with respect to redshift affected by the SMBH mass. 

\begin{figure}[t] \centering
        \includegraphics[width=8cm]{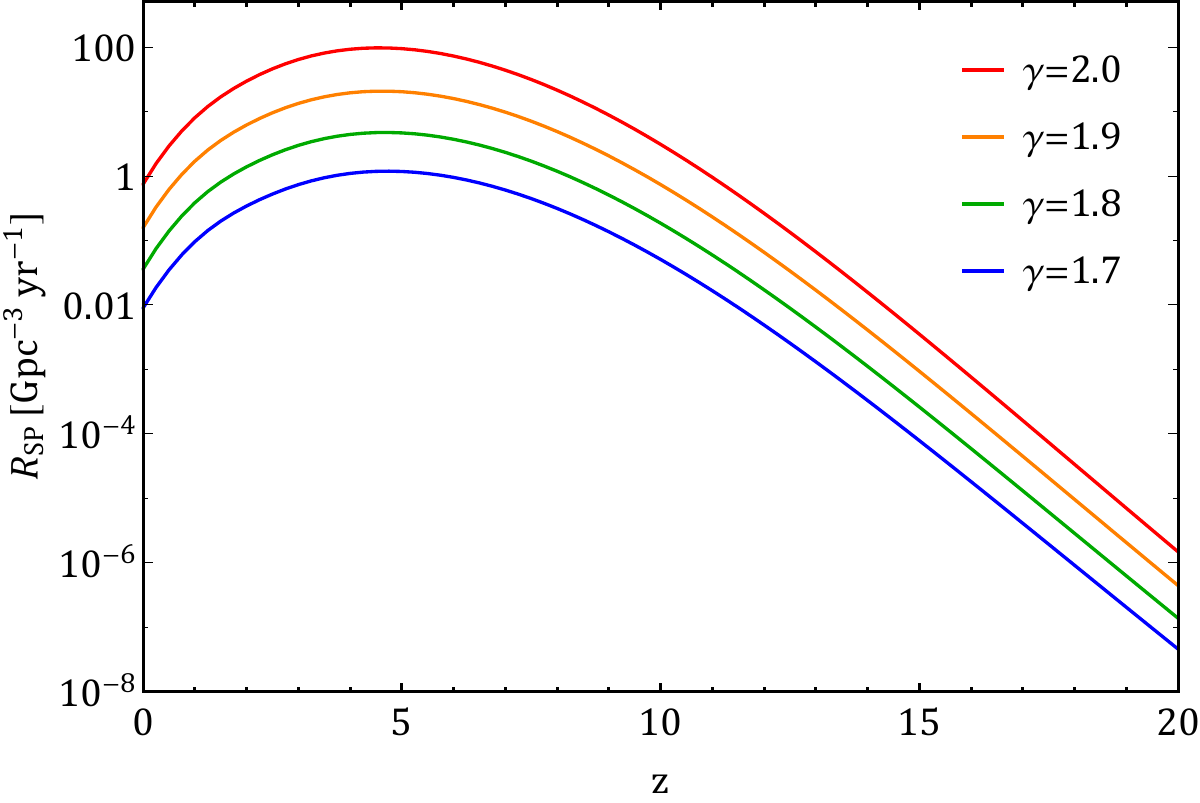}	
	\caption{ 
            Redshift evolution of PBH merger rate in DM spikes for different profiles of power index $\gamma$. 
            Here, we consider $M_\mathrm{PBH} = 30 \, M_\odot$, $f_\mathrm{PBH} = 10^{-3}$ and SMBH masses ranging from $M_{\rm SMBH,min} = 10^5 \, M_\odot$ to $M_{\rm SMBH,max} = 10^9 \, M_\odot$.
	}
        \label{fig:spike_redshift}
\end{figure}

Importantly, as shown in Fig.~\ref{fig:spike_redshift}, peaks in $\dd n/\dd M_\mathrm{SMBH}$ for different SMBH masses result in novel features in the redshift evolution of PBH merger rates associated with DM spikes.
We observe that their overall contribution to the PBH merger rate in Eq.~\eqref{eq:total_merger_spike} results in a distinctive novel peak in PBH merger rate that appears around redshift $z \sim 5$, with similar qualitative behavior found for different considered profiles of DM spikes. 

\begin{figure*}[t] \centering
         \includegraphics[width=8cm]{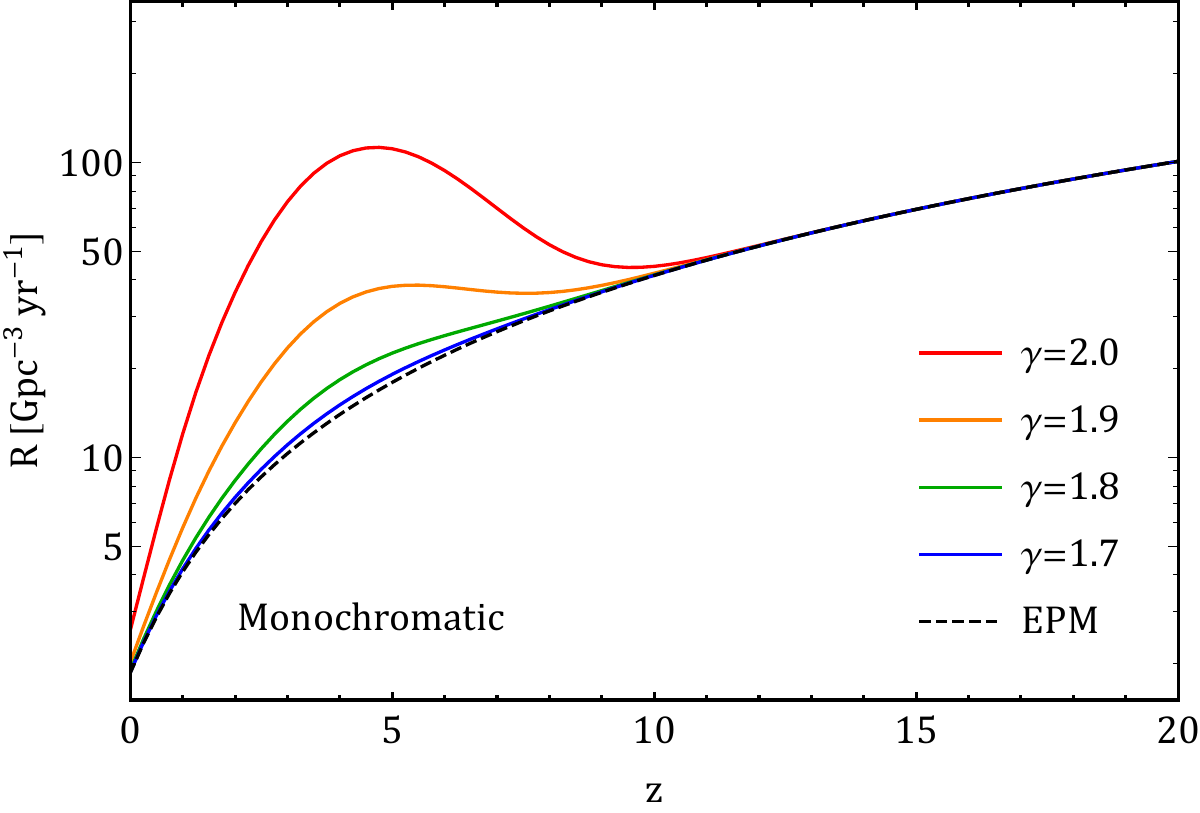} 
         \includegraphics[width=8cm]{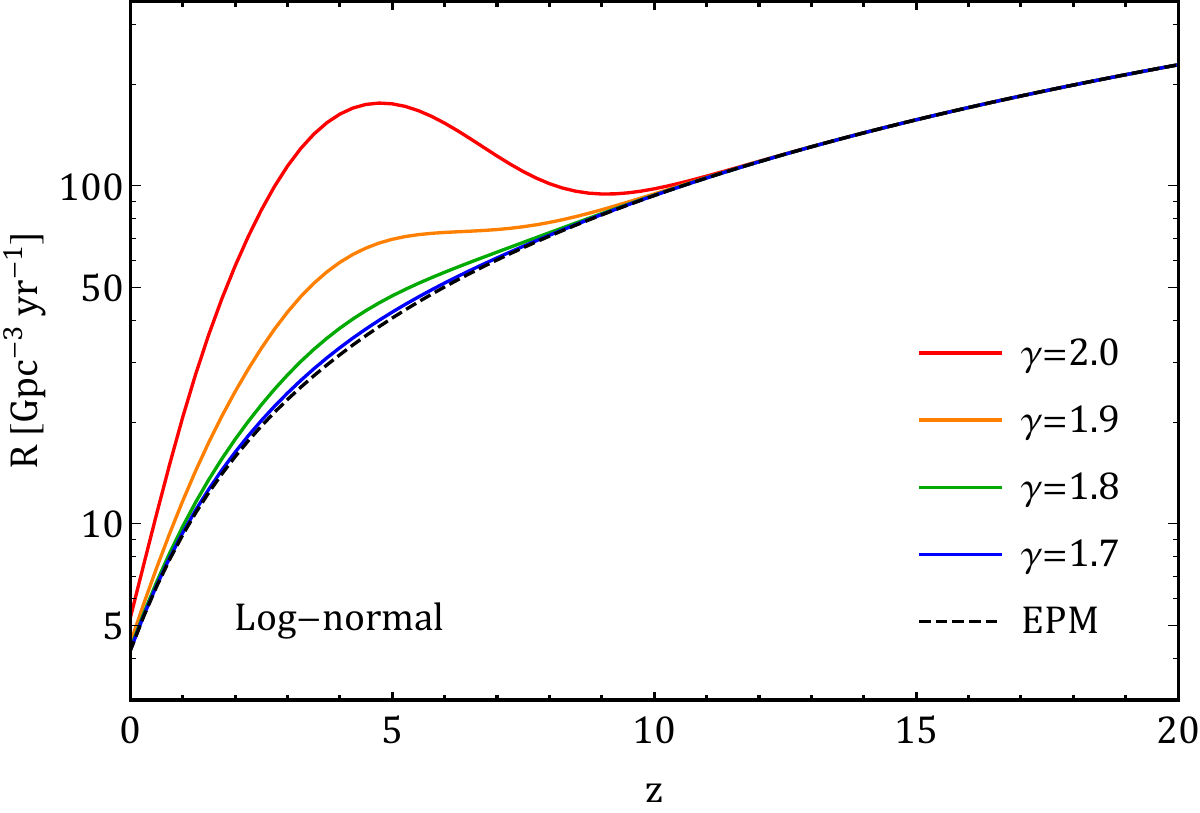} 
	\includegraphics[width=8cm]{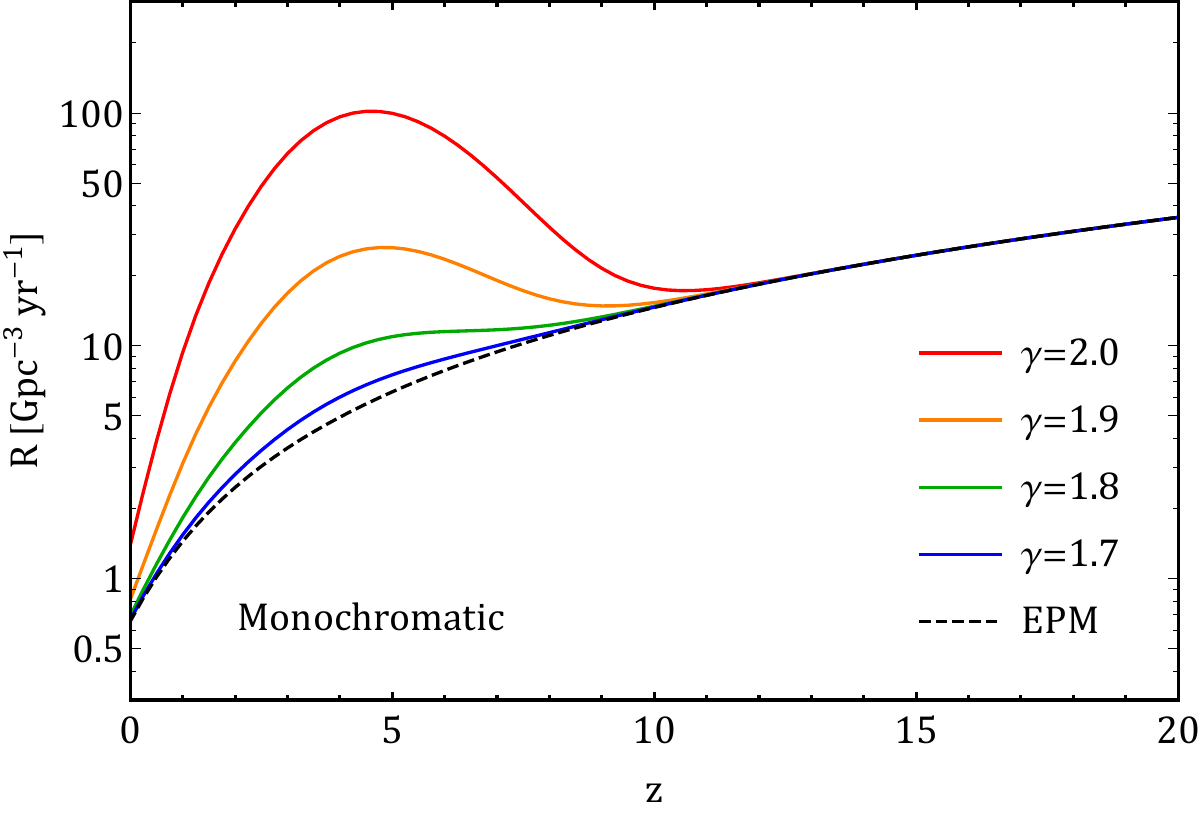}
        \includegraphics[width=8cm]{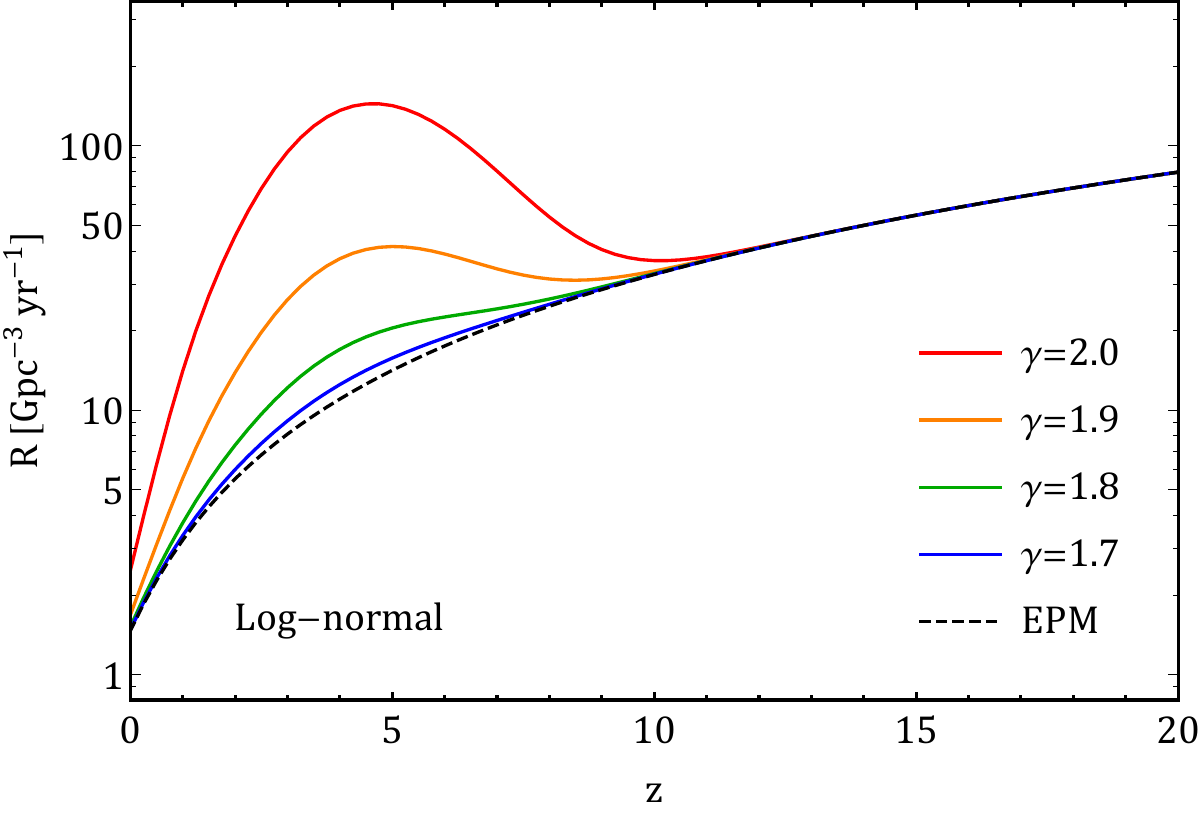}
	\caption{
	[Top Left] Redshift evolution of total PBH merger rate considering monochromatic PBH mass function with $M_{\rm PBH} = 30 \, M_\odot $. [Bottom Left] Redshift evolution of total PBH merger rate considering monochromatic PBH mass function with $M_{\rm PBH} = 100 \, M_\odot $. [Top Right] Redshift evolution of total PBH merger rate considering log-normal PBH mass function with $M_c = 30 \, M_\odot$ and $\sigma = 1$. [Bottom Right] Redshift evolution of total PBH merger rate considering log-normal PBH mass function with $M_c = 100 \, M_\odot$ and $\sigma = 1$. The solid curves are the sum of the contributions from EPM and PBH mergers in DM spikes considering different spike profile index $\gamma$, while the dashed curves represent EPM contributions only. We consider  $f_\mathrm{PBH} = 10^{-3}$ and SMBH masses in the range from $M_\mathrm{SMBH,min} = 10^5 \, M_\odot$ to $M_\mathrm{SMBH,max} = 10^9 \, M_\odot$.
	}
        \label{fig:redshift_modification}
\end{figure*}

To assess the impact of the PBH merger rate redshift evolution in DM spikes on the total PBH merger rate\footnote{PBH merger rates can also be affected by initial conditions, such as clustering~(e.g.~\citet{Young:2019gfc}). We do not discuss these effects here.}, we combine all the relevant contributions, including those from EPM, DM spikes and DM halos. Among these contributions to the overall PBH merger rates, contributions from DM halos are typically expected to be subdominant to that of EPM and DM spikes by several orders~\citep{Bird:2016dcv, Ali-Haimoud:2017rtz, Raidal:2017mfl, Franciolini:2022ewd}. Focusing on the dominant  EPM and DM spike contributions, the total PBH merger rate can be approximated as 
\begin{align}\label{eq:total_merger_approx}
    R_\mathrm{tot} \simeq R_\mathrm{EPM} + R_{\rm SP}~,
\end{align}
where the DM spike contribution $R_{\rm SP}$ is given by Eq.~\eqref{eq:total_merger_spike} and the contribution from EPM can be found by integrating Eq.~\eqref{eq:background_merger} over all PBH masses, considering extended PBH mass-distribution, 
\begin{align}
    R_\mathrm{EPM} = \iint \frac{\dd R}{\dd m_1 \dd m_2} \dd m_1 \dd m_2~.
\end{align}
In Fig.~\ref{fig:redshift_modification} we illustrate the effects of DM spike contributions to the total PBH merger rate of Eq.~\eqref{eq:total_merger_approx} as a function of redshift for monochromatic and log-normal PBH mass-functions, respectively.

We observe in Fig.~\ref{fig:redshift_modification}  novel peaked features in  the redshift evolution of the total PBH merger rate around $z \sim 5$, stemming from PBH DM spike mergers. The peaks originate from the underlying SMBH distribution, with their shapes being dependent on the redshift evolution of $\dd n/\dd M_\mathrm{SMBH}$. As can be seen from Eq.~\eqref{eq:SMBH_massfun}, redshift evolution of $\dd n/\dd M_\mathrm{SMBH}$ is determined by several factors. In particular, the peaked features around $z \sim 5$ result from a larger matter density at higher redshifts, a larger number density of smaller SMBHs (e.g. $ M_{\rm SMBH} \sim 10^5 M_\odot $) compared with the heavier SMBHs, as well as higher SMBH formation rate around $ z\sim 5 $ for the less massive SMBHs.  

The amplitude of the peaks around $z \sim 5$ in PBH merger rate redshift evolution depends on the PBH density in DM spikes and the shape of the peaks depends on the redshift evolution of DM spike profile and SMBH mass distribution. 
With an increased DM spike profile index $\gamma$ the amplitude of the peaks is seen to be enlarged due to a higher contributing PBH number density. 
Further, we find that peaked features in total PBH merger rates become more pronounced compared to EPMs for more massive PBHs. This can be understood from noting that while PBH merger rates in DM spikes are independent of PBH masses since $M_{\rm PBH}$ dependence cancels between Eqs.~\eqref{eq:merger_in_halo} and \eqref{eq:cross_section},
the EPM contributions on the other hand decrease as PBH mass increases, as shown in Eq.~\eqref{eq:background_merger}. 

Comparing PBH merger rates of monochromatic and log-normal mass-functions from Fig.~\ref{fig:redshift_modification} ,
we observe similar peaked behavior. The peaked features in case of extended log-normal distribution effectively average the peaked behavior observed in monochromatic distribution case. Further, as we have discussed above, the peaked features in PBH merger rate with log-normal mass distribution of $M_c = 100 \, M_\odot$ is more pronounced than that with $M_c = 30 \, M_\odot$, due to a larger EPM suppression for heavier PBH masses.

With the development of GW detectors, such as next-generation ground based detector Einstein Telescope~\citep{Punturo:2010zz} and space based detector LISA~\citep{bender1998lisa}, their high sensitivity would allow observations of GW signals to high redshifts. This well positions the GW experiments to probe peaked features in PBH merger rate evolution.

\subsection{Generalized spike profile}
\label{app:generalized_profiles}

Thus far we have considered PBH mergers assuming DM spike profile given by Eq.~\eqref{eq:dm_spike}. Full relativistic calculations~\citep{Sadeghian:2013laa,Speeney:2022ryg} suggest a modified spike profile compared to Newtonian treatment~\citep{Gondolo:1999ef}. Further, spike profile can be modified in case of rapidly rotating Kerr black holes~\citep{Ferrer:2017xwm}. 
While detailed analyses of this is beyond the scope of present work, we can estimate the effects of different spike profiles on PBH merger rates by considering a generalized form 
\begin{align}
    \rho_{\rm sp} = \rho_{\rm R} \left( 1-\alpha \frac{r_s}{r} \right)^k \left( \frac{r_{\rm sp}}{r} \right)^{\gamma_{\rm sp}}~, 
\end{align}
where $ \alpha $ and $ k $ are phenomenological parameters. 
This can well account for relativistic DM spike description.

Starting from Eq.~\eqref{eq:merger_in_spike}, we can separate the PBH merger contributions from the spike profile by considering dependencies from cross-section of Eq.~\eqref{eq:cross_section} and $v_{\rm rel}(r)$, resulting in generalized DM spike PBH merger rate of
\begin{align}
    N_{\rm sp, gen} = A_{s} \int_{\alpha r_s}^{r_{\rm sp}} 
    \left( 1-\alpha \frac{r_s}{r} \right)^{2k} \left( \frac{r_{\rm sp}}{r} \right)^{2\gamma_{\rm sp}- \frac{39}{14}} \dd r~,
\end{align}
where $A_s$ is normalization factor that is fixed by specifying other input factors such as PBH mass function. 
Taking $ \beta \equiv 2\gamma_{\rm sp} -53/14 $, which is positive for the regime of interest with $ \gamma \in [1,2] $, and changing the variables $ t = (r/r_s)^{-\beta} $ this simplifies to 
\begin{align} \label{eq:genspikemod}
    N_{\rm sp, gen} = A_s \frac{1}{\beta} \int_{(r_{\rm sp}/r_s)^{-\beta}}^{\alpha^{-\beta}} \left( 1-\alpha \, t^{1/\beta} \right)^{2k} \dd t ~. 
\end{align}
In the limit of vanishing $r_s/r_{\rm sp} \simeq 0 $, the integration of Eq.~\eqref{eq:genspikemod} can be performed exactly that in terms of Gamma functions $\Gamma$ yields
\begin{align}
    N_{\rm sp, gen} (\alpha, \beta, k) \simeq A_s \frac{1}{\alpha^{\beta} \beta} \frac{\Gamma(1+2k)\Gamma(1+\beta)}{\Gamma(1+2k+\beta)} ~. 
\end{align}
This allows for simple comparison of PBH merger rates stemming from distinct DM spike profiles.

\begin{figure}[t] \centering
        \includegraphics[width=8cm]{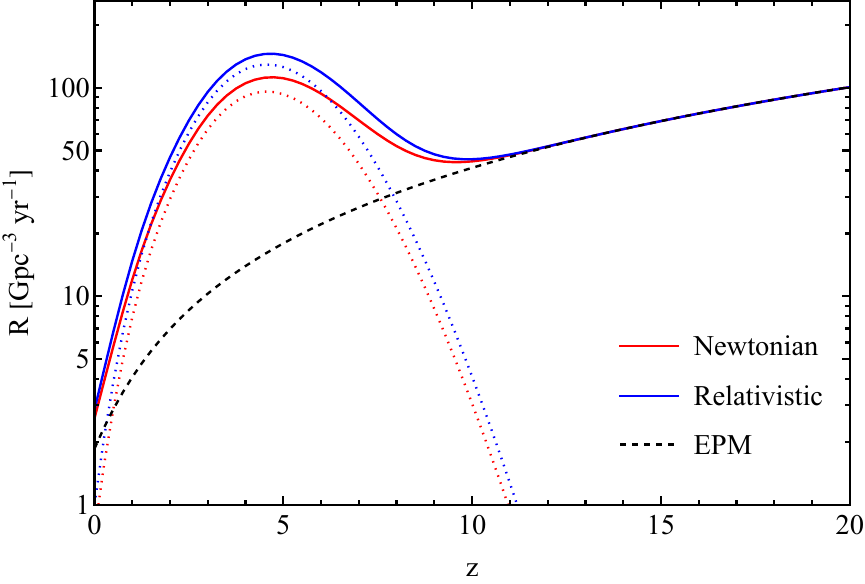}
	\caption{
Comparison between the redshift evolution of PBH merger rates considering Newtonian (red line) and relativistic (blue line) DM spike analyses. We consider the spike power index of $\gamma = 2$, PBHs of mass $M_\mathrm{PBH} = 30 \, M_\odot$ and $f_\mathrm{PBH} = 10^{-3}$.
	}
        \label{fig:relativistic}
\end{figure}

In~Fig.~\ref{fig:relativistic} we illustrate the PBH merger rate redshift evolution in DM spikes comparing Newtonian and relativistic treatments. Considering spike power index of $\gamma = 2$ and thus $ \beta= 17/14$, the Newtonian profile of Ref.~\citet{Gondolo:1999ef} employed in our analysis corresponds to $ \alpha =4 $ and $ k=3$. On the other hand, DM spike profile from relativistic approach of Ref.~\citet{Sadeghian:2013laa} can be well approximated by $ \alpha=2 $ and $ k=5 $. 
Hence, the difference in PBH merger rates between the two is
\begin{align}
    \frac{N_{\rm sp, gen}(2, 17/14, 5) - N_{\rm sp, gen}(4, 17/14, 3) }{N_{\rm sp, gen}(4, 17/14, 3)} \simeq \frac{1}{3}~, 
\end{align}
implying that Newtonian DM spike calculations generally underestimate the merger rate by $\sim1/3$. Thus, our order of magnitude estimates are not significantly affected by these considerations.

\subsection{Non-spherical halo formation}

\begin{figure}[t] \centering
        \includegraphics[width=8cm]{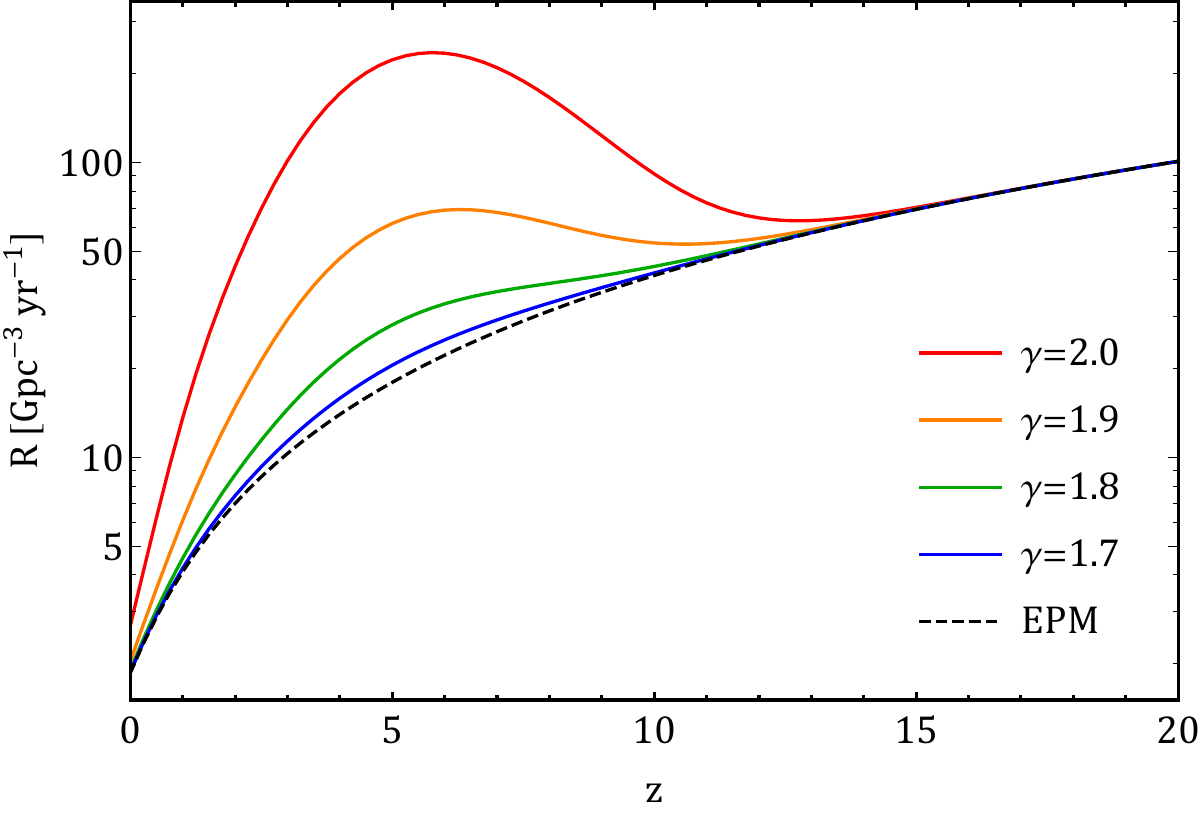}
	\includegraphics[width=8cm]{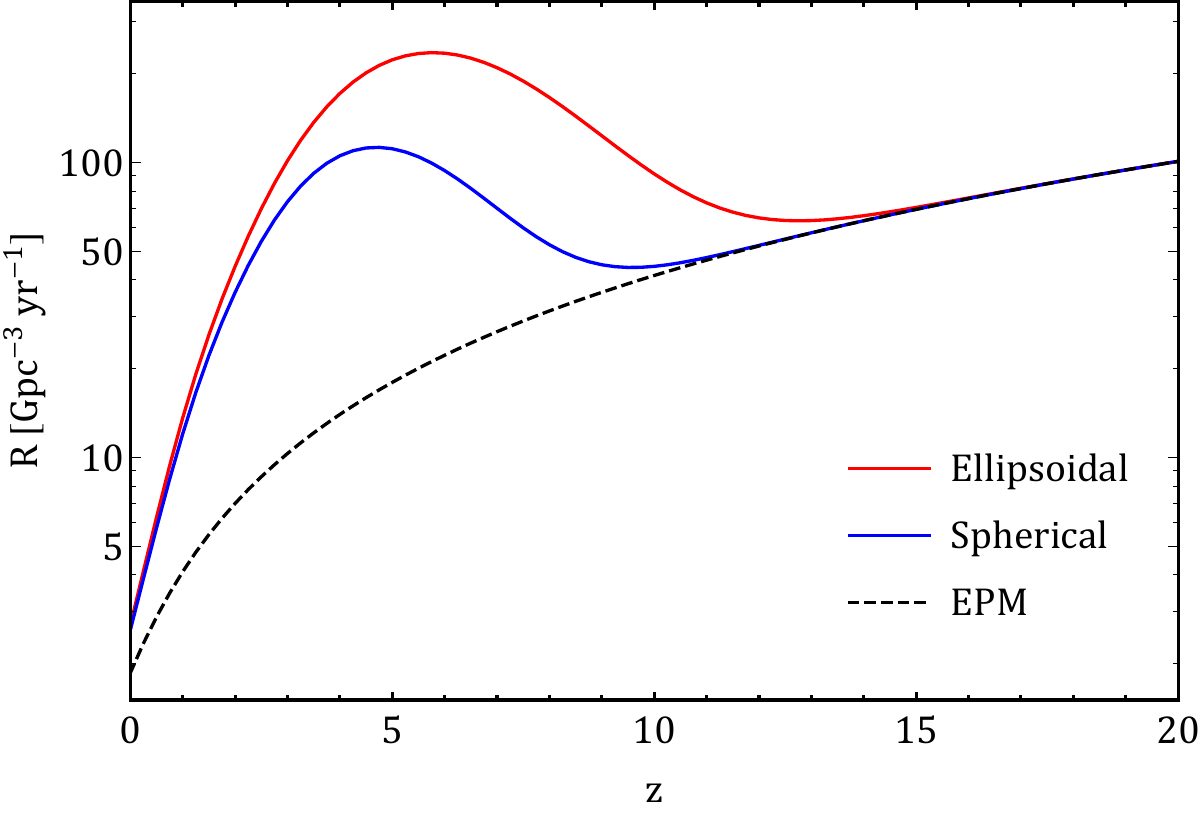}
	\caption{
	[Top] Redshift evolution of the total PBH merger rate with different values of DM spike profile power $\gamma$ considering ellipsoidal-collapse of DM halos. [Bottom] Comparison between the redshift evolution of PBH merger rates in spherical-collapse model and ellipsoidal-collapse model, assuming DM spike profile power index of $\gamma = 2$. Here, we consider $M_\mathrm{PBH} = 30 \, M_\odot$ and $f_\mathrm{PBH} = 10^{-3}$.
	}
        \label{fig:ellipsoidal}
\end{figure}

Throughout we have considered for simplicity the DM halo formation via spherical collapse. However, we can quantify effects on PBH merger rate considering ellipsoidal collapse.  
As shown in earlier studies, ellipsoidal collapse can effectively enhance the merger rate of PBH binaries compared to spherical collapse~\citep{Fakhry:2020plg, Fakhry:2023ggw}. 
Ellipsoidal collapse can be appropriately described by the Sheth \& Tormen (ST) mass distribution~\citep{Sheth:1999mn} as
\begin{align}
    f_\mathrm{ST}(\sigma) = F \sqrt{\frac{2 a}{\pi}} \left[1 + \left(\frac{\sigma^2}{a \delta_\mathrm{c}^2}\right)^p\right] \frac{\delta_\mathrm{c}}{\sigma} \exp \left(- \frac{a \delta_\mathrm{c}^2}{2 \sigma^2}\right) ~,
\end{align}
where $F = 0.322$, $a = 0.707$ and $p = 0.3$. 
$\delta_\mathrm{c} = 1.686$ are parameters for DM halo formation~\citep{Lukic:2007fc}. 
In upper panel of Fig.~\ref{fig:ellipsoidal} we display evolution of PBH merger rates for different DM spike profiles considering ellipsoidal DM halo collapse.
Compared to spherical collapse, we can observe 
in lower panel of Fig.~\ref{fig:ellipsoidal}
that 
ellipsoidal collapse of DM halos can result in a larger peaked feature  around $z \sim 6$. This could be further distinguished with GW observations and detailed analysis of the peaked  features in PBH merger rate evolution can help deepen our understanding of the halo formation history and dynamics.

In our analysis we have focused on standard cold DM halo formation. However, different halo mass functions can be expected for other scenarios such as fuzzy ultralight DM~\citep{Schive:2015kza, Nakama:2020vtw}. This can leave an imprint on peaked features in PBH merger rate evolution that we found. We leave detailed investigation of this for future work.

\subsection{Star formation rate evolution}
\label{app:sfr}

Peaked features in redshift evolution can also be observed in SFR. As we discuss, this can also be attributed to DM halo mass-function evolution.

The SFR can be expressed as~\citep{Tacchella:2018qny}
\begin{align}
    \mathrm{SFR}(M_\mathrm{vir}, z) = \epsilon(M_\mathrm{vir}) f_b \frac{\dd \widetilde{M}_\mathrm{vir}}{\dd t}(M_\mathrm{vir}, z)~,
\end{align}
where $\epsilon(M_\mathrm{vir})$ is the star formation efficiency, which can be estimated as
\begin{align}
    \epsilon(M_\mathrm{vir}) = 2 \epsilon_0 \left[\left(\frac{M_\mathrm{vir}}{M_\mathrm{c}}\right)^{-\beta} + \left(\frac{M_\mathrm{vir}}{M_\mathrm{c}}\right)^{\gamma} \right]^{-1}
\end{align}
and $f_b = \Omega_b/\Omega_m = 0.167$ is the energy density fraction of baryonic matter density in the total matter. We consider $(\epsilon_0, M_c/M_\odot, \beta, \gamma) = (0.26, 7.10 \times 10^{10}, 1.09, 0.36)$ parameters in star formation efficiency.

Here, $\dd \widetilde{M}_\mathrm{vir}/\dd t$ is the delayed and smoothed accretion rate of DM onto its halo
\begin{align}
    \frac{\dd \widetilde{M}_\mathrm{vir}}{\dd t} = \frac{\dd M_\mathrm{vir}}{\dd t} \frac{t_\mathrm{dyn}}{t_\mathrm{SF}}~,
\end{align}
with halo mass growth rate~\citep{2010MNRAS.406.2267F}
\begin{align}\nonumber
    \frac{d M_\mathrm{vir}}{d t} =&~ 46.1 \, M_\odot \, \mathrm{yr}^{-1} \left(\frac{M_\mathrm{vir}}{10^{12} \, M_\odot}\right)^{1.1} (1 + 1.11 z) \\ 
    &\times \sqrt{\Omega_m (1+z)^3 + \Omega_\Lambda}~.
\end{align}
The dynamical time $t_{\rm dyn}$ takes into account delays due to dynamical as well as dissipative effects within halo and given by
\begin{align}
    t_\mathrm{dyn} = \left(\frac{3 \pi}{32 G \rho_\mathrm{crit}}\right)^{1/2} \sim 0.1 t_H~,
\end{align}
where $t_H$ is the Hubble time.
Finally, $t_\mathrm{SF}$ is average timescale to quantify the cumulative star formation processes, which we consider as $3.5 \times 10^9 \, \mathrm{yrs}$. For other parameters, we take those of standard $\Lambda$CDM cosmology.

\begin{figure}[t] \centering
        \includegraphics[width=8cm]{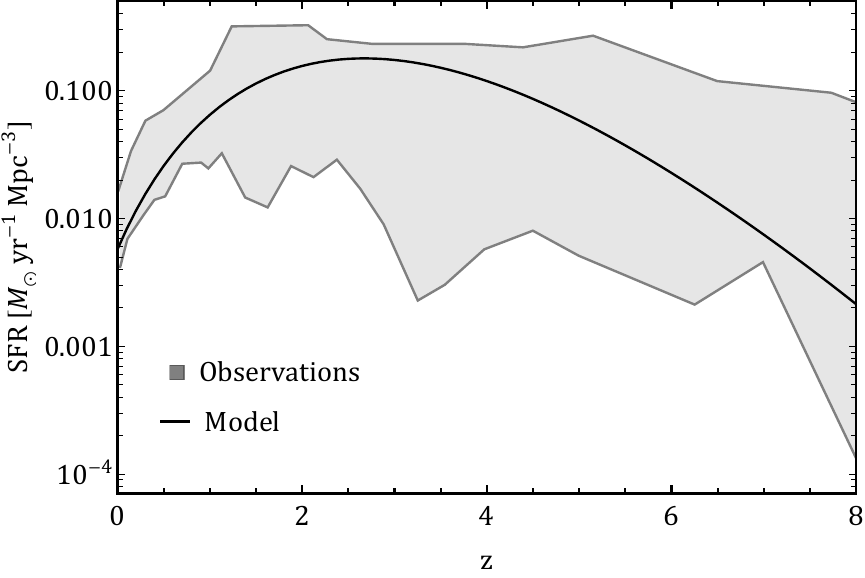}
	\caption{
Calculated SFR at different redshifts based on DM halo mass function from Eq.~\eqref{eq:sfrtot}. Observational multi-messenger bounds \citep{Gruppioni:2020vue}, including IR \citep{Gruppioni:2013jna}, Sub-mm \citep{Gruppioni:2020vue}, UV \citep{Salim:2007is, 2010A&A...523A..74V,  2009ApJ...705..936B, 2011MNRAS.413.2570R, 2012A&A...539A..31C}, GRB \citep{2009ApJ...705L.104K}, SFH passive \citep{2019MNRAS.490.3309M}, Radio \citep{Smolcic:2008br, Dunne:2008bs, 2011ApJ...730...61K} are also displayed.
	}
        \label{fig:SFR}
\end{figure}

The total SFR can be calculated by integrating the SFR for each halo mass following mass function of Eq.~\eqref{eq:halofun}
\begin{align} \label{eq:sfrtot}
    \mathrm{SFR_{tot}}(z) = \int_{M_\mathrm{vir, min}}^{M_\mathrm{vir, max}}\mathrm{SFR}(M_\mathrm{vir}, z) \frac{\dd n}{\dd M_\mathrm{vir}} (z) \, \dd M_\mathrm{vir}~,
\end{align}
where we integrate over DM halo mass range from $M_\mathrm{vir, min} = 10^9 \, M_\odot$ to $M_\mathrm{vir, max} = 10^{15} \, M_\odot$. 
In Fig.~\ref{fig:SFR} we depict our SFR model calculation from Eq.~\eqref{eq:sfrtot} together with various multi-messenger observations  as summarized in Ref.~\citet{Gruppioni:2020vue} and find good agreement within uncertainties.

From Fig.~\ref{fig:SFR} we observe peaked structure in redshift evolution of SFR around $z \sim $~few, in analogy with peaked structure in redshift evolution of PBH merger rates that we found. This can be understood as follows. From our model we can identify that this depends on two factors. One is the SFR redshift evolution and the other is halo mass function ${\rm d}n/{\rm d}M_\mathrm{vir}$ at different redshifts. 
Compared with peaked structure in PBH merger rate evolution, SFR reaches its maximal value at somewhat lower redshifts of around $z \sim 2.5$. This is caused by the different redshift dependence between $\mathrm{SFR}(M_\mathrm{vir}, z)$ in total SFR evolution and $\dd M_\mathrm{vir}/\dd M_\mathrm{SMBH}$ of Eq.~\eqref{eq:SMBH_massfun} that enters PBH merger rate evolution computation. 
Peaked structure in SFR and PBH mergers are thus both associated with peaked structure found in DM halo mass function redshift evolution.

\section{Dark matter spike evolution}
\label{sec:evolutionspike}

In our analysis thus far we has focused on the PBH merger rate in stable DM spikes after formation, considering that density profile of spikes is not evolving. However, competing effects can affect these results. Among them is two-body relaxation that
suppresses the DM spike merger contributions. Another significant effect is loss-cone repopulation, which enhances the merger contributions. Our simplified analysis highlights the need for comprehensive understanding of these complex effects with additional dedicated studies and simulations.

\subsection{Relaxation}\label{app:two_body_relaxation}

  Two-body relaxation can play a significant role in dynamics of astrophysical N-body systems~\citep{2000chun.proc...76M, 2014ApJ...786..121S, 2013pss5.book..923S}. Since relaxation timescale and lengthscale are determined by the number and density of celestial objects,
in spikes high DM density is flattened within a short relaxation timescale and the PBH merger rate can be effectively suppressed by several orders of magnitude~\citep{Nishikawa:2017chy}.
This merger rate suppression can significantly dampen the peaked features around $z \sim 5$ in the evolution of PBH merger rates. When PBH density is sufficiently small, the probability of two-body encounters between PBHs is decreased and thus two-body relaxation effects are also weakened.

DM spike two-body relaxation depends on the relaxation time scale $t_\mathrm{relax}$, which is related to the relaxation length scale $r_\mathrm{relax}$ as  \citep{2013pss5.book..923S}
\begin{align}\label{eq:relaxation_timescale}
    t_\mathrm{relax} = \frac{v_\mathrm{rel}^3(r_\mathrm{relax})}{8 \pi G^2 M_\mathrm{PBH} f_\mathrm{PBH} \rho_\mathrm{sp}(r_\mathrm{relax}) \log(b_\mathrm{max}/b_\mathrm{min})}~,
\end{align}
where $b_\mathrm{min}$ and $b_\mathrm{max}$ are the impact parameters corresponding to the Schwarzschild radius of SMBH and radius of the DM spike, respectively. As before,
for $v_{\rm rel}$ we use Eq.~\eqref{eq:vrel} that depends on SMBH mass $M_\mathrm{SMBH}$. From larger $M_\mathrm{SMBH}$ and smaller $M_\mathrm{PBH}$ with smaller $f_\mathrm{PBH}$ reduce two-body relaxation effects. 

At different redshifts $z$ the relaxation timescale $t_\mathrm{relax}$ of Eq.~\eqref{eq:relaxation_timescale} can be estimated as Hubble time difference $t_{\rm relax}(z) = t_H(z) - t_{H}(z_{\rm form})$ between redshift $z$ and DM halo formation redshift $z_\mathrm{form}$, which we set  $z_\mathrm{form} \simeq 20$. Then $r_{\rm relax}$ can be numerically solved from the right-hand side of Eq.~\eqref{eq:relaxation_timescale}, and gives a redshift evolution of $r_\mathrm{relax}$ due to two-body relaxation.
We then can construct the density profile of DM spike including relaxation effects for different redshifts. A smoothed core with a density $\rho_\mathrm{sp}(r_\mathrm{relax})$ forms in the region $r < r_\mathrm{relax}$ where relaxation time $t_{\rm relax} < t_H$. The remaining DM density profile is taken to be redistributed as the profile of DM spike in the region of $r_\mathrm{relax} < r < \tilde{r}_\mathrm{sp}$, where $\tilde{r}_\mathrm{sp}$ is determined by considering where the total mass of the initial DM spike equates the total mass of the relaxed DM spike. Then, the relaxed density profile of DM spike can be described  as
\begin{equation}\label{eq:relax_profile}
    \rho_\mathrm{sp,relax}(r) = 
    \begin{cases}
        \rho_\mathrm{sp}(r_\mathrm{relax}), ~~~{\rm for }~ 4 r_\mathrm{s} < r < r_\mathrm{relax}\\
        \rho_\mathrm{sp}(r),~~~~~~~~ {\rm for }~ r_\mathrm{relax} < r < \tilde{r}_\mathrm{sp} 
    \end{cases}
\end{equation}
After determining the relaxed density profile, the mass-changing rate due to relaxation $\Gamma_{\rm TBR} = \dd M_{\rm TBR}/\dd t$ can be evaluated by calculating the total relaxed mass at different redshifts as
\begin{align} \label{eq:mtbr}
    M_{\rm TBR}(z) = \int_{4 r_s}^{r_{\rm relax}(z)} 4 \pi \rho_{\rm sp}(z)  r^2 \dd r~,
\end{align}
then taking derivative with respective to time.

\begin{figure}[t] \centering
        \includegraphics[width=8cm]{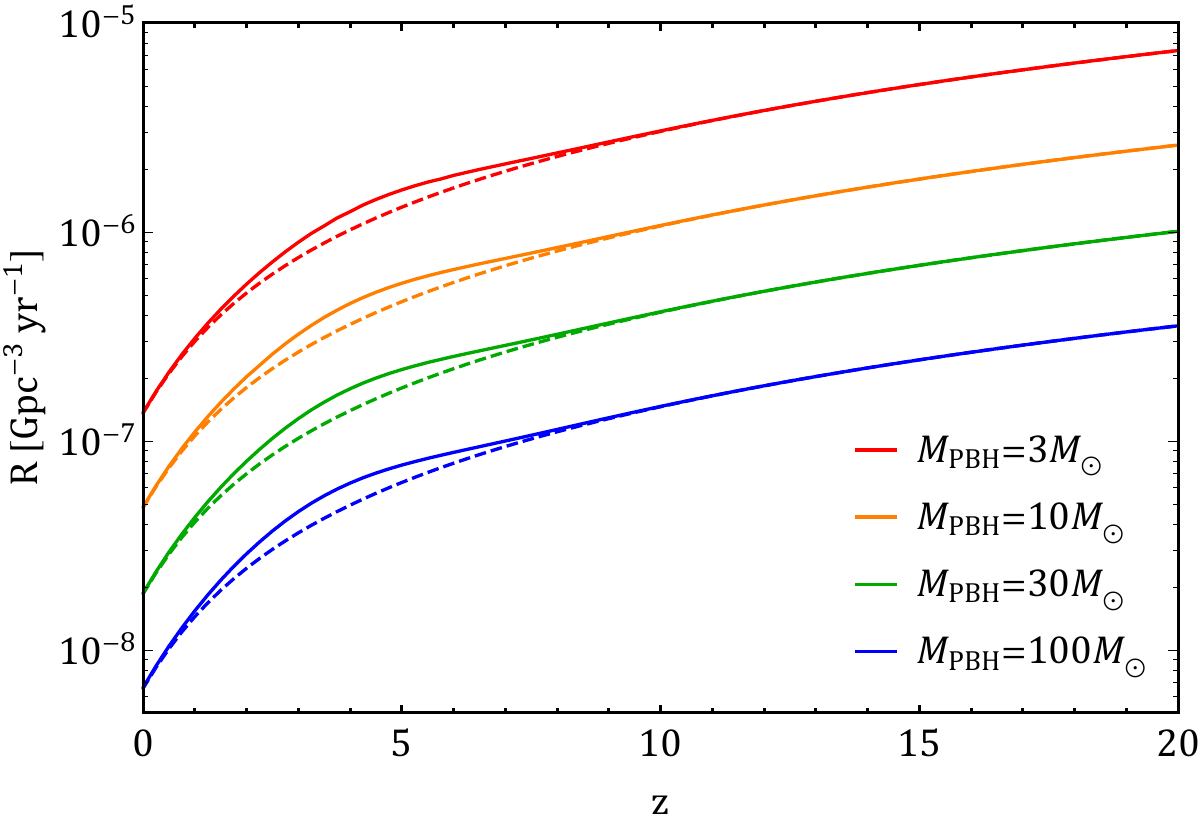}
	\includegraphics[width=8cm]{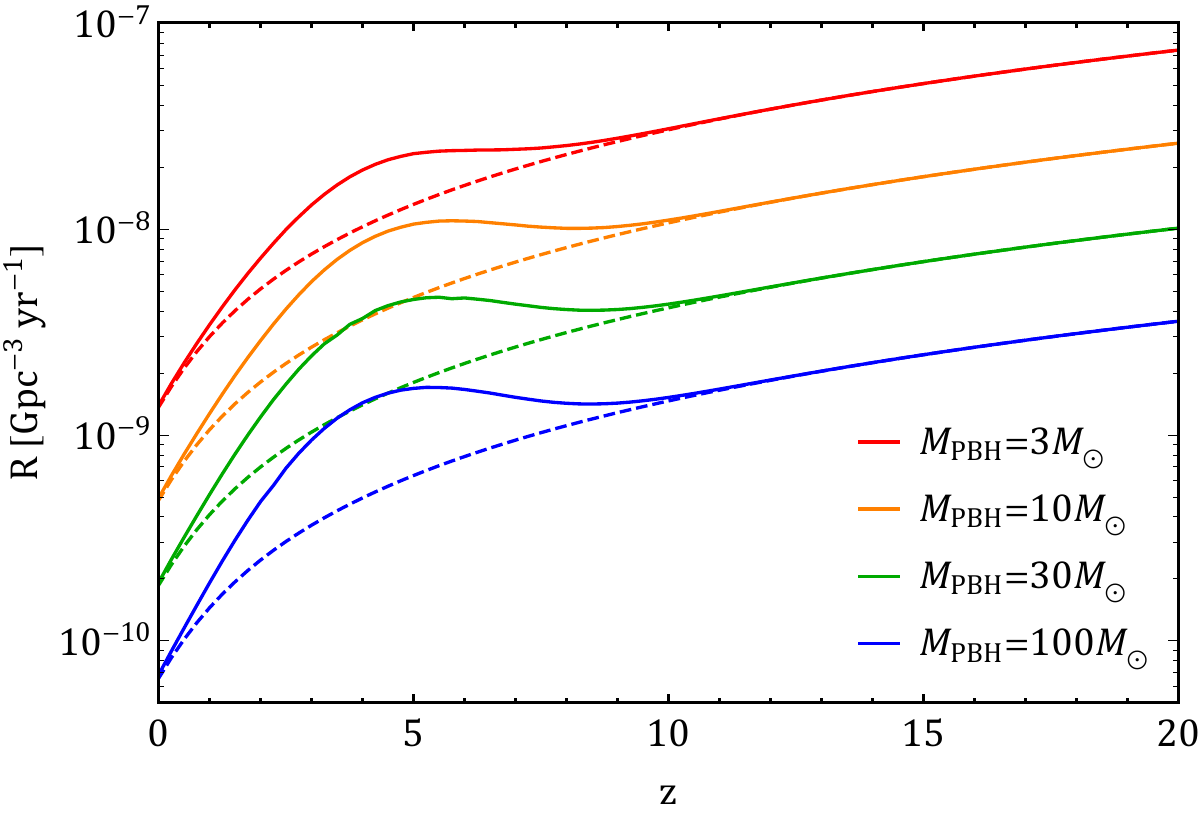}
	\caption{
	Redshift evolution of total PBH merger rate including contributions from EPMs and DM spike with two-body relaxation effects for different PBH masses. We consider $f_\mathrm{PBH} = 10^{-7}$ [Top] and $f_\mathrm{PBH} = 10^{-8}$ [Bottom], DM spike profile power index $\gamma = 2$ and contributions from SMBHs in the mass range from $M_\mathrm{SMBH, min} = 10^5 \, M_\odot$ to $M_\mathrm{SMBH, max} = 10^{10} \, M_\odot$.
	}
        \label{fig:relaxation}
\end{figure}

In Fig.~\ref{fig:relaxation} we display combined PBH merger rate evolution with respect to redshift from Eq.~\eqref{eq:total_merger_approx} considering relaxed DM spike density profile of Eq.~\eqref{eq:relax_profile} in Eq.~\eqref{eq:merger_in_spike}. 
We find that two-body relaxation effects can significantly affect peaked structure in PBH merger rate evolution found earlier, smoothing it out for larger values of $f_{\rm PBH}$. As effects of two-body relaxation become suppressed for smaller $f_{\rm PBH}$ values, peaked structure in PBH merger rate evolution persists and becomes unaffected by them when  $f_\mathrm{PBH} \lesssim 10^{-7}$.
Since the total PBH merger rate includes contributions from EPM, we find that PBH mass does not significantly impact qualitative behavior of the peaked structure due to increase in EPM contributions while the DM spike relaxation decreases in Eq.~\eqref{eq:relaxation_timescale}.

\subsection{Loss-cone refilling} \label{app:loss_cone_refilling}

The loss-cone consists of orbits interacting with the central SMBH, which can be refilled through gravitational encounters in DM halo and thus increase PBH density (see e.g. Ref.~\citet{2013CQGra..30x4005M} for review). Studies on loss-cone refilling highlight its complexity~\citep{2013CQGra..30x4005M, Milosavljevic:2002ht, Merritt:2004fr, 2013ApJ...774...87V, 2014ApJ...785..163V, Avramov:2020aye}. 
Estimates find stellar capture rates of around $\sim (10^{-4} - 10^{-6}) M_\odot/ \mathrm{yr} $ for SMBHs of mass $M_\mathrm{SMBH}$ in the range $\sim (10^{6} - 10^{10})  M_\odot$~\citep{2013ApJ...774...87V}, which in case of PBHs could significantly repopulate the DM spike over galactic timescales of $\sim 10^{10}$ yrs. N-body simulations find increased refilling rates for triaxial and axisymmetric halos~\citep{Gualandris2016CollisionlessLR}. 

We consider a simplified effective treatment for loss-cone refilling to gain insights into its qualitative behavior. 
We assume that refilling of DM would gradually reconstruct the DM spike density profile that is smoothed by relaxation Eq.~\eqref{eq:relax_profile}.
Following Ref.~\citet{2013ApJ...774...87V}, which analyzed axisymmetric galaxies, we consider the loss-cone refilling rate  as
\begin{align}\label{eq:refilling_rate}
    \Gamma_{\rm LCR} = \frac{\dd M}{\dd t} \simeq f_\mathrm{PBH} \frac{M_\mathrm{PBH}}{M_\mathrm{SMBH}} \frac{\sigma^3}{G}~,
\end{align}
where $\sigma$ is velocity dispersion of DM halo that can be calculated from $M_\mathrm{SMBH} - \sigma$ relation in Eq.~\eqref{eq:mass_sigma}. Note that in our simplified treatment we do not include here redshift dependence besides that of $\sigma(z)$ stemming from Eq.~\eqref{eq:mass_sigma}.

\begin{figure}[t] \centering
        \includegraphics[width=8cm]{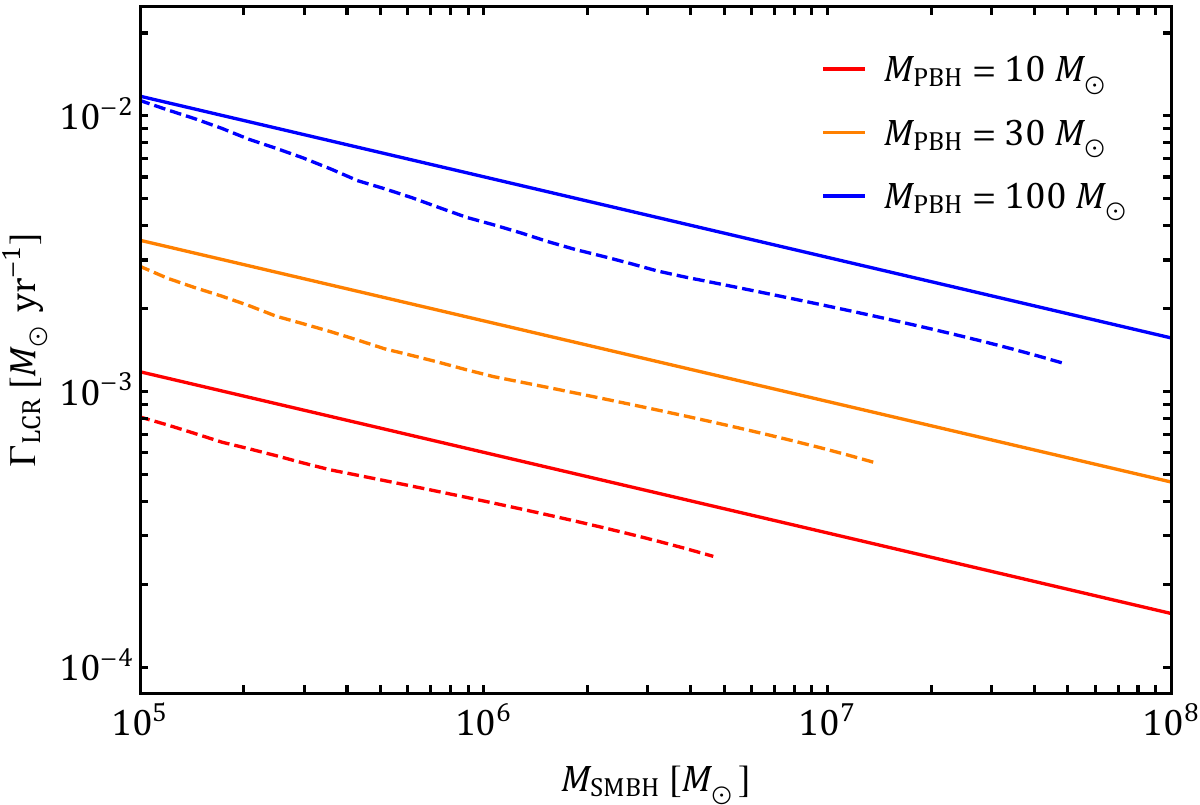}
        \includegraphics[width=8cm]{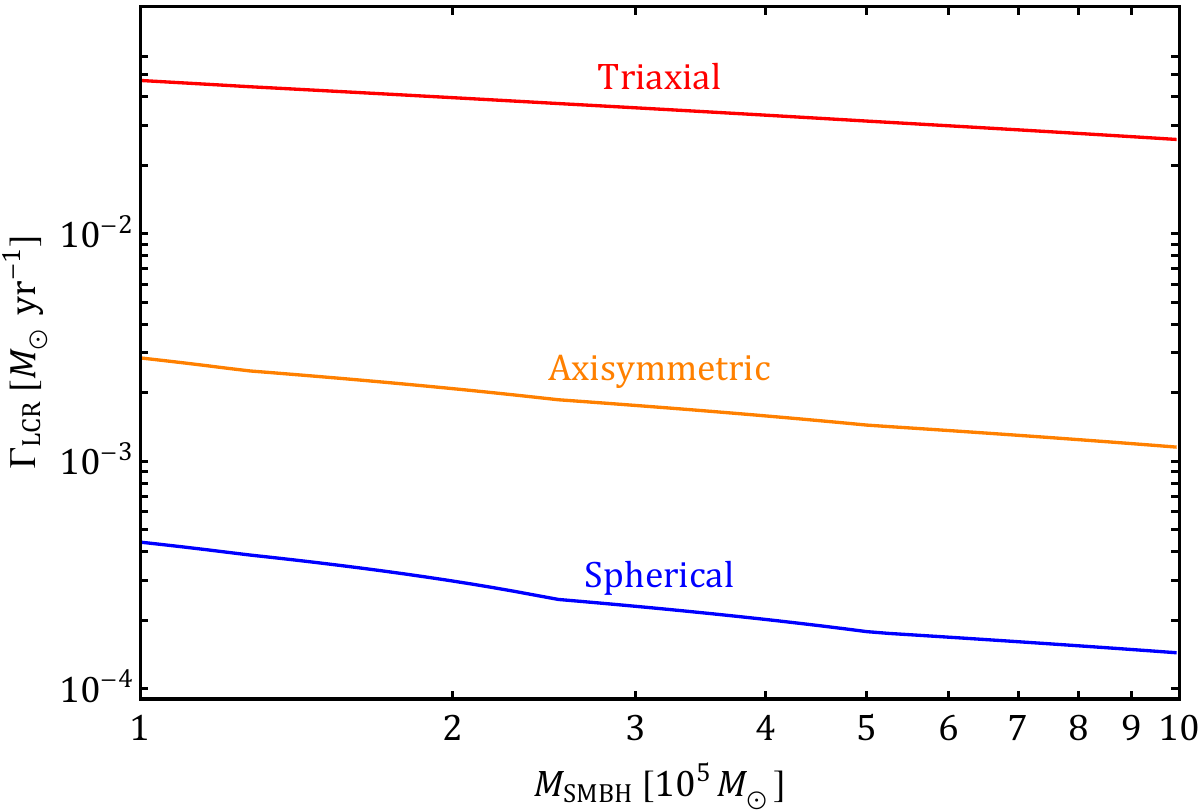}
	\caption{
	[Top] Relation between loss-cone refilling rate and SMBH mass for different PBH masses considering analytical estimates (solid curves) and N-body simulations (dashed), in case of axisymmetric galaxies. [Bottom] The relation between the loss-cone refilling rate and SMBH mass in triaxial, axisymmetric, and spherical galaxies, considering PBH mass of $30 \, M_\odot$.
	}
        \label{fig:refilling_rate}
\end{figure}

To compare with simulation results in~\citet{Gualandris2016CollisionlessLR}, we calculate the loss-cone refilling rate by averaging the total mass of refilling simulated particles over the relaxation timescale as follows  
\begin{align}\label{eq:refilling_rate_simu}
    \frac{\dd M}{\dd t} = \frac{M_\mathrm{sim} R_J}{t_\mathrm{relax,lc}} = \frac{N M_\mathrm{PBH} R_J}{t_\mathrm{relax,lc}}~,
\end{align}
where $M_\mathrm{sim} = N M_\mathrm{PBH}$ is the total mass of simulated particles and $N$ is their total number, $R_J$ is the fraction of refilling particles in the total number of simulated particles. Here, $t_\mathrm{relax,lc}$ is the relaxation timescale for the loss-cone refilling, which can be estimated as \citep{binney2011galactic}
\begin{align} \label{eq:relaxation_timescale_2}
    t_\mathrm{relax,lc} \simeq \frac{N}{8 \log N} \frac{1}{\sqrt{G M_\mathrm{PBH} n}}~,
\end{align}
where $n$ is the number density of simulated particles. Then we calculate the dependence of loss-cone refilling rate on PBH mass, SMBH mass, and their behaviors in various types of DM halos.
In the upper panel of Fig.~\ref{fig:refilling_rate}, we compare loss-cone refilling rate for axisymmetric galaxies from Eq.~\eqref{eq:refilling_rate} with N-body simulation results using Eq.~\eqref{eq:refilling_rate_simu} and observe qualitative agreement differing within a factor of few.  

Note that two-body relaxation time in Eq.~\eqref{eq:relaxation_timescale} and Eq.~\eqref{eq:relaxation_timescale_2} corresponds to the same mechanism that flattens DM spike density profile and also is associated with repopulating loss-cone, however these timescales apply in different environments. Namely, relaxation time of Eq.~\eqref{eq:relaxation_timescale} is considered in DM spike, while relaxation time of Eq.~\eqref{eq:relaxation_timescale_2} is associated with DM halo. Starting from a general relaxation timescale of Eq.~\eqref{eq:relaxation_timescale} that we used for DM spike we can also consider it in the context of loss-cone refilling, assuming a relaxation system with size $r_\mathrm{relax}$, $v_\mathrm{vel} \sim \sqrt{GM/r_\mathrm{relax}} \sim \sqrt{G N M_\mathrm{PBH}/r_\mathrm{relax}}$, $b_\mathrm{max} \sim r_\mathrm{relax} \sim (N/n)^{1/3}$ and $b_\mathrm{min} \sim n^{-1/3}$, then Eq.~\eqref{eq:relaxation_timescale} and Eq.~\eqref{eq:relaxation_timescale_2} can be put in the same form.

Using N-body simulation results\footnote{
Different considerations of loss cone angular momentum can lead to variation of results within a factor of few~\citep{Gualandris2016CollisionlessLR}, however their qualitative behavior remains similar.}
of Ref.~\citet{Gualandris2016CollisionlessLR} (see e.g. their Fig. 5) for different types of galaxies we calculate the loss-cone refilling rates as a function of SMBH mass. The results are shown in the lower panel of Fig.~\ref{fig:refilling_rate}, depicting that loss-cone refilling is more efficient in triaxial and axisymmetric halos rather than spherical. Since the majority of galaxies are not spherical, the role of spherical halos in this process could be expected to be subdominant when the overall galaxy population is appropriately considered. On the other hand, while detailed simulations are lacking, near galactic center DM halo might be naively expected to be more spherical. The loss-cone refilling rate also depends on several other factors. Higher refilling rates result from larger PBH masses and number densities, as well as smaller masses of SMBHs.

\subsection{Combined effects}

To estimate the impact of DM spike evolution effects on PBH merger rate evolution, we consider competing
DM spike two-body relaxation effect and with loss-cone refilling. 
Then we combine two effects.
First we estimate the evolution of the total mass in DM spike, which neglects other possible contributions as follows
\begin{align}
    \Gamma_{\rm SP} \simeq \Gamma_{\rm LCR}  + \Gamma_{\rm TBR}~,
\end{align}
where subscript LCR and TBR denote the loss-cone refilling and two-body relaxation. The mass evolution due to two-body relaxation $\Gamma_{\rm TBR}$ is found from Eq.~\eqref{eq:mtbr}. The loss-cone refilling contribution $\Gamma_{\rm LCR}$ is calculated from Eq.~\eqref{eq:refilling_rate}. Knowing the DM spike mass evolution at different cosmic times and redshifts, we can reconstruct the density profile of DM spike based on Eq.~\eqref{eq:relax_profile}. Then, we calculate the PBH merger rates in DM spike and study their redshift evolution. 

\begin{figure}[t] \centering
        \includegraphics[width=8cm]{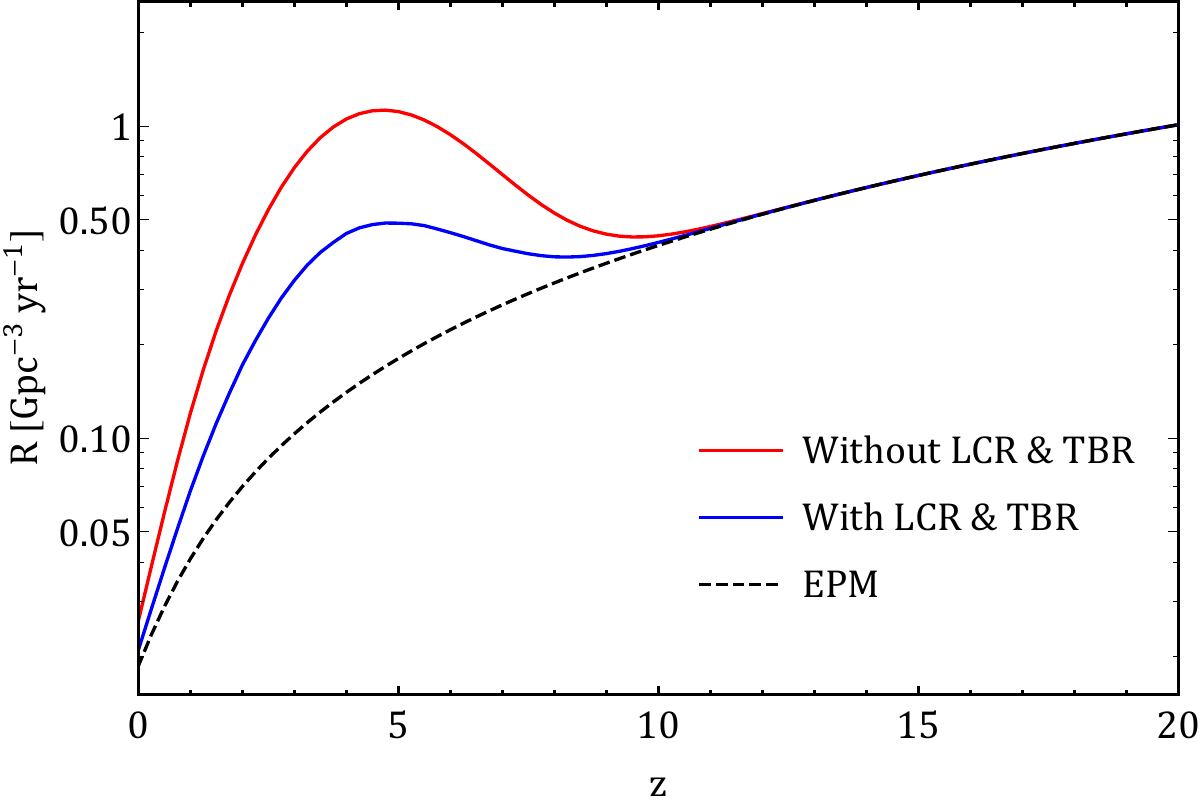}
        \includegraphics[width=8cm]{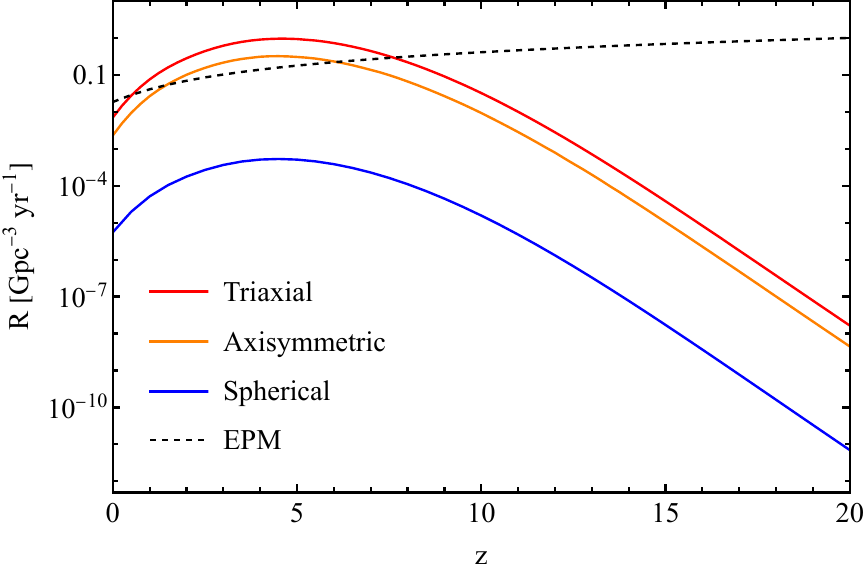}
	\caption{
	[Top] Redshift evolution of total PBH merger rate including DM spike evolution two-body relaxation and loss-cone refilling effects, assuming axisymmetric galaxy. [Bottom]   Redshift evolution of PBH merger rate in DM spike including DM spike evolution two-body relaxation and loss-cone refilling effects in three types of galaxies: triaxial (red line), axisymmetric (orange line), and spherical galaxies (blue line). We consider $M_\mathrm{PBH} = 30 \, M_\odot$, $ \gamma = 2$, and $f_\mathrm{PBH} = 10^{-4}$.
	}
        \label{fig:total_evol_effect}
\end{figure}

In Fig.~\ref{fig:total_evol_effect} we display the effects on PBH redshift evolution including DM spike evolution two-body relaxation and loss-cone refilling.
This demonstrates that for PBHs with $M_\mathrm{PBH} = 30 \, M_\odot$ and $f_\mathrm{PBH} = 10^{-4}$ in axisymmetric DM halos PBH merger rate evolution results in formation of discernible peaked structure, albeit suppressed compared to the case without spike evolution effects. For larger $f_{\rm PBH}$ we find two-body relaxation dominates the peak behavior, further suppressing it. For significantly smaller $f_{\rm PBH}$, two-body relaxation effects become diminished. Based on the types of contributing DM halos, the PBH merger rate in DM spikes can be quite distinct, where the triaxial DM halos contribute the largest PBH merger rates in DM spikes, while spherical DM halos result in smaller amounts of PBH mergers in DM spikes. 

Contributions from distinct galaxy morphologies~\citep{binney2011galactic} need to be properly accounted for to  comprehensively capture the impact of these complex effects on the total PBH merger rates, which is beyond the scope of this work. The predominant majority of galaxies are not purely spherical (e.g.~\citet{Maccio:2008pcd,2011MNRAS.415L..69J,Rodriguez:2013uea,2019MNRAS.484..476C}) and hence their contributions where peaked features in PBH merger rate evolution are found to be suppressed can be expected to be subdominant. Our findings call for dedicated simulations and analyses of these effects and in distinct populations of galaxies.

\begin{figure*}[t] \centering
        \includegraphics[width=8cm]{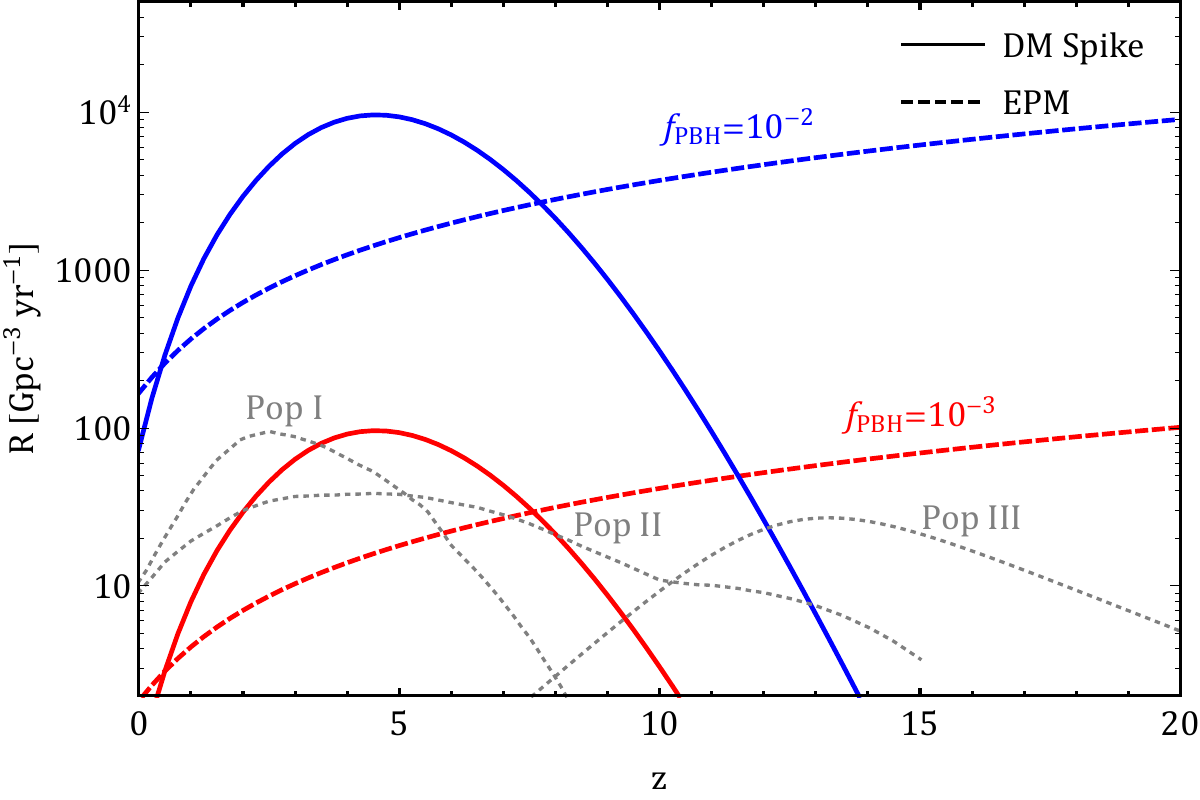}
        \hspace{1cm}
        \includegraphics[width=8cm]{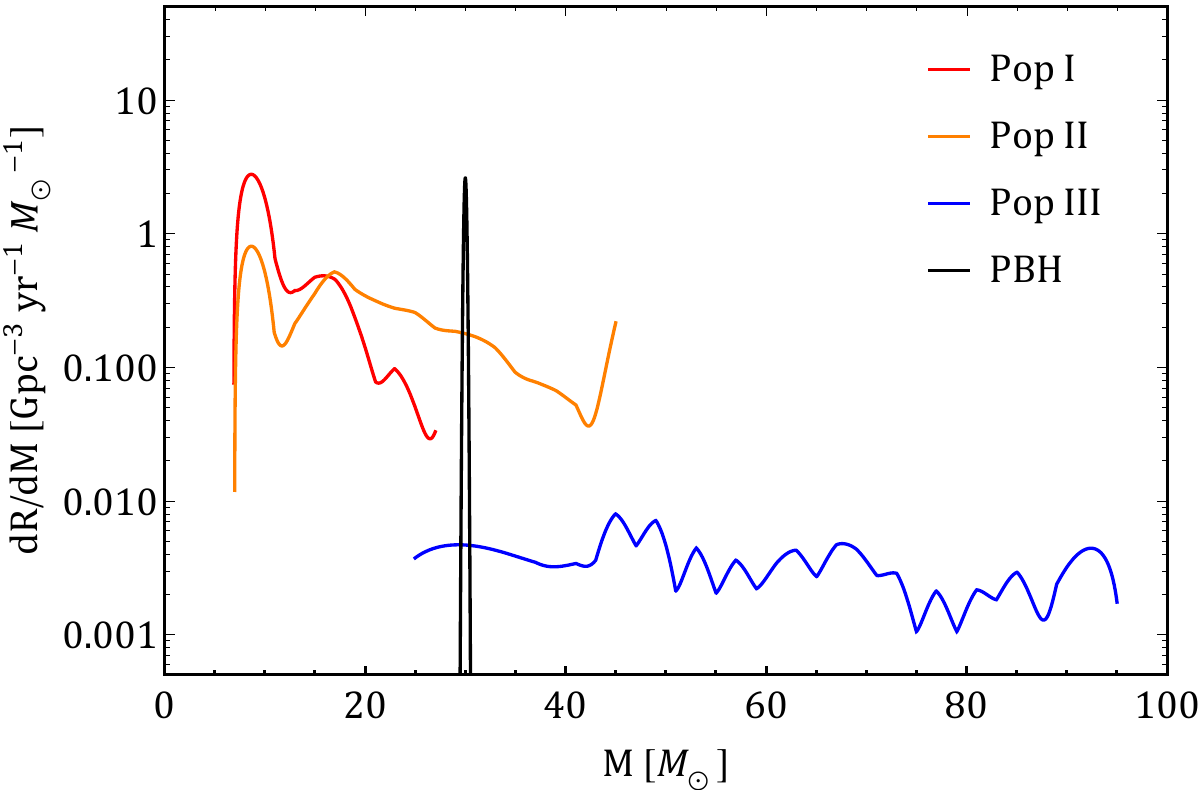}
	\caption{
	[Left] Redshift evolution of merger rates from PBHs and Pop I/II/III BHs. 
        The dotted curves for Pop I/II BHs are taken from Ref.~\citet{Tanikawa:2021qqi} and the dotted curve for Pop III is calculated in Eq.~\eqref{eq:popiii}. 
        The red and blue dashed curves are the PBH merger rates from early PBH binaries with $f_\mathrm{PBH} = 10^{-3}, 10^{-2}$, respectively. 
        The red and blue solid curves are the PBH merger rates from DM spike with $f_\mathrm{PBH} = 10^{-3}, 10^{-2}$, respectively. 
        We set $M_\mathrm{PBH} = 30 \, M_\odot$ and $\gamma = 2$. [Right] The mass distribution of differential merger rates from Pop I/II/III BHs and PBHs at redshift $z = 0$. 
        We set $f_\mathrm{PBH} = 10^{-3}$ and $M_\mathrm{PBH} = 30 \, M_\odot$. 
        The differential merger rates for Pop I/II/III BHs are taken from Ref.~\citet{Tanikawa:2021qqi}
	}
        \label{fig:pop3_total_redshift}
\end{figure*}

\section{Astrophysical and primordial black holes}
\label{sec:astrobh}

For observations, PBH contributions should be combined with that of astrophysical black holes. 
We consider astrophysical black holes that are remnants of three generations of stars, that is Population I/II/III (Pop I/II/III). 
Their merger rates have been extensively studied~(e.g.~\citet{Belczynski:2016ieo, Tanikawa:2021qqi}). 

We account for the merger rate redshift evolution of astrophysical black holes as follows.
For black holes associated with Pop I/II stars we consider population synthesis as described in Ref.~\citet{Tanikawa:2021qqi} and for Pop III star black holes we employ phenomenological model of Ref.~\citet{Ng:2020qpk} in order to calculate their merger rates up to redshift $z \sim 20$. 
The merger rate of Pop III black holes can be parameterized as~\citep{Ng:2020qpk}
\begin{align}\label{eq:popiii}
    R_\mathrm{III}(z) \propto  \frac{ e^{a_\mathrm{III}(z - z_\mathrm{III})}}{a_\mathrm{III} + b_\mathrm{III}e^{(a_\mathrm{III} + b_\mathrm{III})(z - z_\mathrm{III})}}~,
\end{align}
where we consider input parameters $(a_\mathrm{III}, b_\mathrm{III}, z_\mathrm{III}) = (0.66, 0.3, 11.6)$ as Ref.~\citet{Ng:2022agi} from 
fitting to population synthesis results~\citep{Belczynski:2016ieo}. 
The overall normalization can be found by setting the peak value to be $ \sim 10\% $ of that in Pop I/II black hole merger rates~\citep{Belczynski:2016ieo, Tanikawa:2021qqi}, which gives $R_\mathrm{III}(z_\mathrm{III}) = 20 \, \mathrm{Gpc}^{-3} \, \mathrm{yr}^{-1}$~\citep{Ng:2020qpk, Belczynski:2016obo}.  

In left panel of Fig.~\ref{fig:pop3_total_redshift} we display the resulting redshift evolution of astrophysical Pop I/II/III black hole merger rates overlaid together with PBH merger rates including contributions originating from EPM and DM spikes as discussed earlier.
We observe that evolution of Pop I/II/III astrophysical black hole merger rates and PBH merger rates from DM spikes have maximum peaks at different redshifts. This signifies these contributions can be distinguished and peaks in observed black hole merger rates around $z \sim 5$ can serve as novel probes of concentrated DM spikes in galactic centers.

Characteristics of PBH merger rates from DM spikes allow to further distinguish them from Pop I/II/III astrophysical black holes.
While PBH merger rates are sensitive to PBH abundance $f_{\rm PBH}$ as shown in Fig.~\ref{fig:pop3_total_redshift}, astrophysical black hole merger rates are independent of it.
PBH merger rate increases with PBH abundance and for $f_{\rm PBH} \gtrsim 10^{-3}$ dominates over astrophysical black hole merger rate redshift evolution.

As we display in right panel of Fig.~\ref{fig:pop3_total_redshift} BH mass distributions also differ between PBHs and astrophysical black holes. Here, we consider merger rates of astrophysical Pop I/II/III black holes from Ref.~\citet{Tanikawa:2021qqi} (see also Ref.~\citet{Fryer:1999ht, 2010ApJ...725.1918O} for other black hole mass distribution studies), and the PBH merger rates are calculated assuming a monochromatic distribution around mass $M_\mathrm{PBH} = 30 \, M_\odot$.
We observe that PBH merger rates can be distinguished in mass and also can dominate over astrophysical black holes.

\section{Conclusions}
\label{sec:summary}

With the prevalence of SMBHs at the centers of galaxies, their influence on surrounding DM distribution can significantly affect the merger rates of PBHs that could have formed in the early Universe, contribute to DM abundance and that have been linked to recent GW detections. 

We identify novel peaked structure around redshift $z \sim 5$ in evolution of PBH merger rates stemming from contributions of enhanced DM density spikes around SMBHs in the late Universe. 
The peaked features arise from redshift evolution of SMBH mass function as well as enhancement of PBH merger rates due to increased DM density in spikes. 
We find that merger rate evolution of heavier PBHs exhibits a more pronounced peaked structure due to lower contributions of EPMs. The DM spike profile  characterized by power index $\gamma$ also influences the prominence of merger rate evolution peaks, with $\gamma \gtrsim 1.7$ resulting in more discernible merger rate peaks.
Two-body relaxation and loss-cone refilling can influence the merger rate evolution, potentially suppressing its peaked behavior. However, such effects are subject to significant uncertainties. Our work highlights the need for further investigations and simulations of these effects to improve understanding and for more accurate predictions.
We note that similar peaked structure also appears in the redshift evolution of SFR. 

Contributions from astrophysical Pop I/II/III BHs may also create features at different redshifts, which can in principle be convoluted with those stemming from PBHs. However, unlike astrophysical black holes, PBH contributions increase with $f_\mathrm{PBH}$ making the peaked structure more discernible. Further, differentiating black hole population properties such as mass distribution of PBHs and astrophysical black hole could aid in distinguishing their respective contributions.

In summary, the newly identified peaked structure in PBH merger rate redshift evolution could serve as a probe of DM in galactic centers, with its amplitude and shape providing insights into DM density profiles and SMBH mass distributions. Future GW detectors like Einstein Telescope and LISA, with their improved sensitivity, could further enhance our ability to detect and study these features, offering a window into PBH merger rates and DM distribution in the Universe.

\section*{Acknowledgements}
We thank Ilias Cholis, Tomoya Kinugawa, Misao Sasaki, John Silverman as well as Kavli IPMU cosmology group for discussion. Q.D. was supported by IBS under the project code, IBS-R018-D3. M.H. was supported by IBS under the project code, IBS-R018-D1. V.T. acknowledges support of the JSPS KAKENHI grant No. 23K13109 and World Premier International Research Center Initiative (WPI), MEXT, Japan. This work was performed in part at the Aspen Center for Physics, which is supported
by the National Science Foundation grant PHY-2210452.
 
\bibliographystyle{aasjournal}
\bibliography{references}

\end{document}